\documentclass[3p, 11pt]{elsarticle}
\usepackage{graphicx,amscd,amsmath,amssymb,verbatim}
\usepackage{amsfonts,epsfig}
\usepackage{mathptmx}
\usepackage{setspace}
\usepackage{epsfig}
\usepackage{url}
\usepackage{multirow}
\usepackage{color}
\usepackage{subfigure}

\begin{document}

\journal{Elsevier}

\begin{frontmatter}

\title{Deterministic and stochastic aspects of the stability in  an inverted pendulum 
under a generalized parametric excitation}

\author{Roberto da Silva, Sandra D. Prado, Debora E. Peretti} 

\address{Institute of Physics, Federal University of Rio Grande do Sul,
Av. Bento Gon\c{c}alves, 9500, Porto Alegre, 91501-970, RS, Brazil
{\normalsize{E-mail:rdasilva@if.ufrgs.br}}}


\begin{abstract}

In this paper, we explore the stability of an inverted pendulum under
a generalized parametric excitation described by a superposition of $N$
cosines with  different amplitudes and frequencies, based on a simple
stability condition that does not require any use of Lyapunov exponent, for 
example. Our analysis is separated in 3 different cases: $N=1$, $N=2$, and $N$ 
very large. Our results were obtained via numerical simulations by fourth-order Runge
Kutta integration of the non-linear equations. We also calculate the effective
potential also for $N>2$. We show then that numerical integrations recover a wider 
region of stability that are not captured by the (approximated) analytical
method. We also analyze stochastic stabilization here: firstly, we look the
effects of external noise in the stability diagram by enlarging the
variance, and secondly, when $N$ is large, we rescale the amplitude by showing
that the diagrams for time survival of the inverted pendulum resembles the exact
case for $N=1$. Finally, we find numerically the optimal number of cosines corresponding
to the maximal survival probability of the pendulum.

\end{abstract}

\end{frontmatter}


\setlength{\baselineskip}{0.7cm}

\section{Introduction}

The inverted pendulum, more precisely its stabilization mechanisms deserve a
lot of attention from several correlated areas, including Physics,
Mathematics, Biology, (see for example an interesting review \cite%
{Ibrahim2006}). However, the applications goes beyond, including the study
of excitation effects in Ocean Structures \cite{Ocean1998}, inverted
pendulum robots \cite{Kim2005} and many others.

Induced stability is a solved problem known since 1908 \cite{Stephenson1908}%
, but it was in the 1950s, with Kapiza \cite{Kaptiza1951}, that this kind of
stability was studied for the inverted pendulum system. Experimental results
were obtained in the 1960s (see for example \cite{Bogdanoff1965}) and even
nowadays this problem still remains interesting \cite{Chinas2013}. The
possible excitations/perturbations which may be capable of stabilizing an
inverted pendulum, even for some time, have a large number of details which
are not completely understood yet. Therefore, this apparent simple system,
has a complexity that can be considered a challenging problem since some
important stabilization problems in Engineering, as for instance, the
stability of robot arms, the stability of populations in biology \cite%
{Ibrahim2006}, or even the stabilization of photons deviation in Cosmology 
\cite{Prado2004} require similar stabilization mechanisms.

\begin{figure}[th]
\begin{center}
\includegraphics[width=0.4\textwidth,height=0.3%
\textheight]{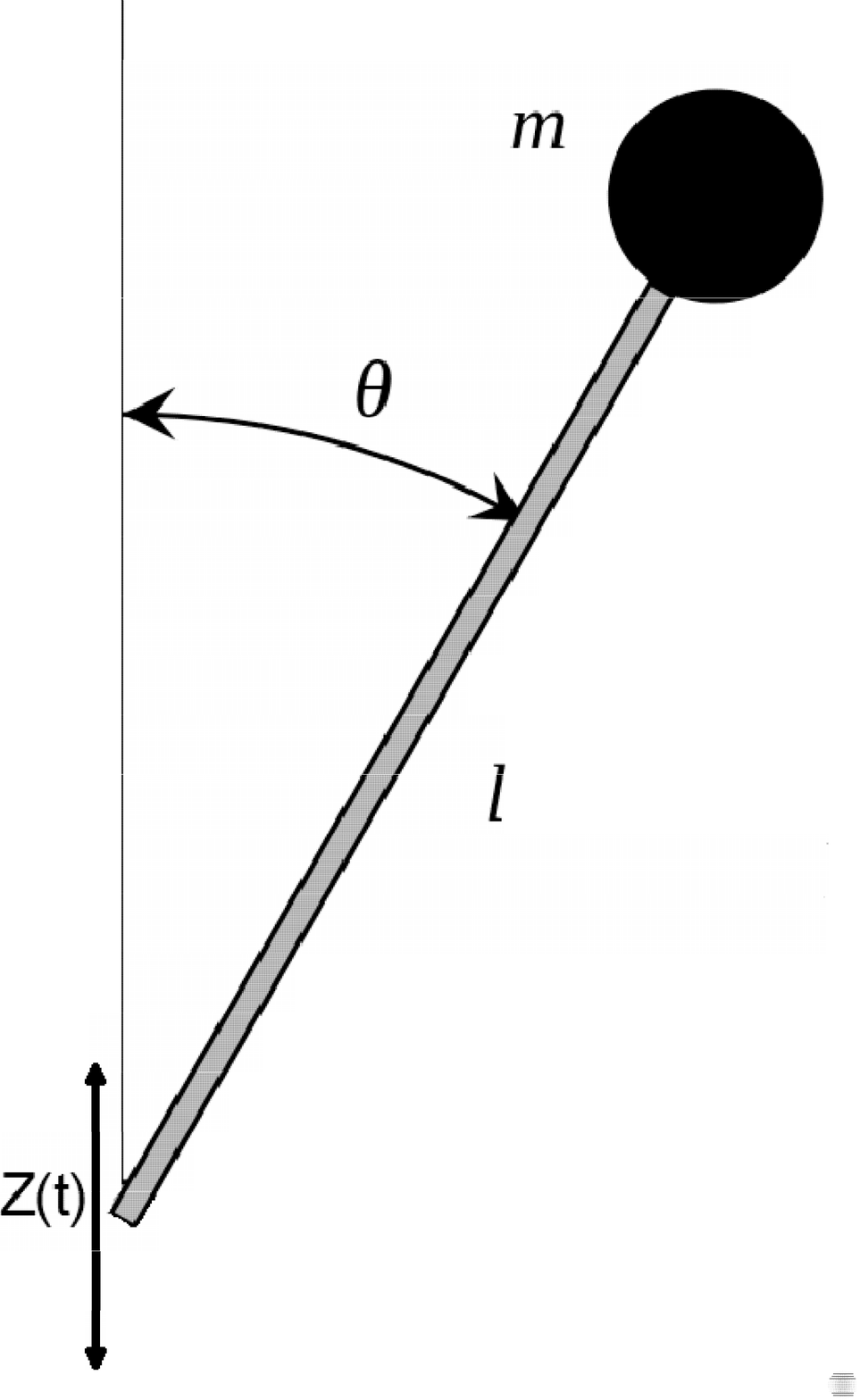}
\end{center}
\caption{Inverted pendulum under a support excitation $z(t)$. }
\label{Fig:Inverted_pendulum}
\end{figure}

An inverted pendulum (showin in Fig. \ref{Fig:Inverted_pendulum}) free of
external forces, is unstable and the punctual mass $m$ attached to the rigid
massless rod will tend to oscillate around to the stable equilibrium
position ($\theta =180^{0}$), which corresponds to the usual pendulum
problem).

In order to keep the pendulum upright, $\cos \theta >0$, the frictionless
hinge that attaches the rod to the suspension point point, must vertically
accelerated. Let us denote such acceleration by $a(t)=\ddot{z}(t)$, where $%
z(t)$ is a time-dependent excitation that controls the height of the
pendulum suspension point $P$.

The Lagrangian of this problem can easily be written as:%
\begin{equation*}
\mathcal{L}(\theta ,\dot{\theta})=\frac{1}{2}ml^{2}\dot{\theta}^{2}-ml\dot{z}%
\dot{\theta}\sin \theta -mgl\cos \theta -mgz(t)\text{.}
\end{equation*}%
It is worth emphasizing that all our results can be directly extended to an
equivalent physical pendulum making simple associations. If we additionally
consider a external excitation $\phi (t)$ we can derive the motion equation

\begin{equation}
\ddot{\theta}(t)=\frac{g}{l}\left( 1+\frac{1}{g}\ddot{z}(t)\right) \sin
\theta +\phi (t)  \label{complete}
\end{equation}%
from the Lagrange equations.

Now, in order to generalize, we write $z(t)=z_{\text{det}}(t)+z_{\text{rand}%
}(t)$ and $\phi (t)=\phi _{\text{det}}(t)+\phi _{\text{rand}}(t)$ where
subscripts \textquotedblleft det\textquotedblright\ and \textquotedblleft
rand\textquotedblright\ denote the deterministic and stochastic
time-dependent parts of the excitations/perturbations respectively. In this
work, we focus our analysis in two important situations: $z_{\text{rand}%
}(t)=0$ and $z_{\text{det}}(t)=$ $\sum_{j=1}^{n}A_{j}\cos (\omega _{j}t)$,
i.e., parametric sinusoidal excitation in the basis and $\phi _{\text{det}%
}(t)=0$ and $\phi _{\text{rand}}(t)=N(\mu ,\sigma )$ that denotes an
external gaussian noise with mean $\mu $ and standard deviation $\sigma $.
The more appropriate choice here is $\mu =0$, given that we are mainly
interested in the parametric stabilization of the inverted pendulum or in
what we can call its survival time $\tau $, which means the maximum time the
pendulum remains upright or, mathematically, the time up to condition $\cos
\theta <0$ is satisfied. The condition $\mu \neq 0$ leads to a natural
biased motion, which is not interesting here. In this work we consider to
analyze how the stability condition is broken as function of the external
noise dispersion $\sigma $.

Several authors \cite{Ibrahim2006,Butkov2001} have explored the case $N=1$
using small angles approximation known as the effective potential method.
However, even the well known case for one cosine, $N=1$, deserves, in our
opinion, some attention and alternative analysis. Therefore, firstly, we
checked the literature based on numerical integration methods in order to
verify the validity of previous results in the small oscillations regime and
the initial conditions dependence, basically, the initial angle dependence.
Particularly we also use these numerical integrations to observe the
breaking of the stability region predicted by perturbative analysis and by
parametric resonance (see Butikov \cite{Butkov2001}) when the external
noise is turned on (i.e., $\phi \neq 0$). In this case we calculate the
pendulum survival probability over different time evolutions (i.e, different
evolutions corresponding to different seeds). A detailed connection between $%
N=1$ and $N\rightarrow \infty $ is also explored when the pendulum amplitude
grows linearly with time and is also rescaled with the number of cosines
set in the parametric excitation.

In the second part of this work we analyze the superposition of two cosines, 
$N=2$, which as far as we are concerned has not been explored yet. This case
is interesting since the period of the composition is not always known so
that the perturbative analysis is not able to describe the inverted pendulum
stability regions correctly. Given such a problem, we analyze numerically
the interesting case where the amplitudes are fixed and the frequencies: $%
\omega _{1}\ $and $\omega _{2}$ are varied. The analytical results obtained
via the effective potential method are compared with numerical simulations.
We also analyze the effects of an analysis of the stability breakdown
considering different external noise dispersion $\sigma $ in the numerical
diagrams. Deviations from small oscillations behavior were considered in our
analysis.

Finally, for $N>2$, we look for optimization problems on the stability
probability considering an ensemble of frequencies and amplitudes. The most
remarkable detail in our analysis is the fact a simple choice for the
stability criterion as, $\cos \theta >0$, overcomes more laborious methods
such as Liapunov exponent or other analysis to check stability conditions.
So we organized our papers as it follows: in the next section we describe
key points for a perturbative analysis extending our formulation for
arbitrary $N$, which is known in literature as effective potential method.
In section \ref{Section:Numerical Simulations}, we briefly show our
numerical simulations were developed. Our main results are presented in
section \ref{Section:Results}. Some discussions and conclusions are finally
presented in section \ref{Section:Conclusions}.

\section{Perturbative Analysis}

\label{Section:Perturbative_Analysis}

We start our perturbative analysis by chosing the perturbative function $%
z(t)=\sum_{i=1}^{N}A_{i}\cos (\omega _{i}t)$, so that the Eq. \ref{complete}
becomes 
\begin{equation}
\frac{d^{2}\theta }{dt^{2}}=\frac{g}{l}\left( 1-\frac{1}{g}%
\sum_{i=1}^{N}A_{i}\omega _{i}^{2}\cos (\omega _{i}t)\right) \sin \theta
\label{Eq:effective_1}
\end{equation}%
where we set up $\phi (t)=0$ for all $t$.

The last equation can be written in a more elucidative form as 
\begin{equation}
\overset{\cdot \cdot }{\theta }\ =-\frac{\partial U}{\partial \theta }%
+F(\theta )  \label{Eq:Movement}
\end{equation}%
where $U=\frac{g}{l}\cos \theta $ and $F(\theta )=\frac{\overset{..}{z}(t)}{l%
}\sin \theta =-\frac{\sin \theta }{l}\sum_{i=1}^{N}A_{i}\omega _{i}^{2}\cos
(\omega _{i}t)$. The physical interpretation is straightforward here, given
that $U(\theta )$ is the gravitational potential, while $F(\theta )$ is an
external force due to vibrations at the pendulum suspension point. In what
follows from here, we will adapt the equations obtained by Landau and
Lifchitz \cite{Landau}. A similar analysis was developed by Kapitza (1951) 
\cite{Kaptiza1951}.

We start assuming that the solution of Eq. \ref{Eq:Movement} can be
separated into two aditive components: 
\begin{equation}
\theta (t)=\overline{\theta }(t)+\xi (t)\text{.}
\label{Eq:perturbative_solution}
\end{equation}%
We are considering that the pendulum motion is composed by a non-perturbated
path $\overline{\theta }$ plus a noise $\xi $ composed by multiple
frequencies $\{\omega _{i}\}_{i=1}^{N}$and amplitudes $\{A_{i}\}_{i=1}^{N}$.
Another way to think is that $\overline{\theta }(t)$ has a large amplitude
but slow frequency while $\xi (t)$ has small amplitude but fast frequency.

By the Taylor expansions of $U(\theta )$ and $F(\theta )$ around the slow
path $\overline{\theta }$, we have: 
\begin{eqnarray}
\frac{\partial U}{\partial \theta } &=&\left. \frac{\partial U}{\partial
\theta }\right\vert _{\theta =\overline{\theta }}+\left. \frac{\partial ^{2}U%
}{\partial \theta ^{2}}\right\vert _{\theta =\overline{\theta }}\xi +\frac{1%
}{2}\left. \frac{\partial ^{3}U}{\partial \theta ^{3}}\right\vert _{\theta =%
\overline{\theta }}\xi ^{2}+...  \label{Eq:perturbative_expansions} \\
\frac{\partial F}{\partial \theta } &=&F(\overline{\theta })+\left. \frac{%
\partial F}{\partial \theta }\right\vert _{\theta =\overline{\theta }}\xi
+\left. \frac{\partial ^{2}F}{\partial \theta ^{2}}\right\vert _{\theta =%
\overline{\theta }}\xi ^{2}+...  \notag
\end{eqnarray}

Considering these approximations up to the first order term, we can obtain
from Eq. \ref{Eq:Movement} that 
\begin{equation}
\overset{\cdot \cdot }{\overline{\theta }}+\overset{\cdot \cdot }{\xi }\
=-\left. \frac{\partial U}{\partial \theta }\right\vert _{\theta =\overline{%
\theta }}-\left. \frac{\partial ^{2}U}{\partial \theta ^{2}}\right\vert
_{\theta =\overline{\theta }}\xi +F(\overline{\theta })+\left. \frac{%
\partial F}{\partial \theta }\right\vert _{\theta =\overline{\theta }}\xi
\label{Eq:non-perturbated_oscilating}
\end{equation}

Now it is crucial to consider the nature of motion to distinguish the
important terms in Eq. \ref{Eq:non-perturbated_oscilating}. The only
candidates associated with the perturbative effects on the right side of
this equation are $-\left. \frac{\partial ^{2}U}{\partial \theta ^{2}}%
\right\vert _{\theta =\overline{\theta }}\xi $, $F(\overline{\theta })$, and 
$\left. \frac{\partial F}{\partial \theta }\right\vert _{\theta =\overline{%
\theta }}\xi $. Therefore, given that the terms $-\left. \frac{\partial ^{2}U%
}{\partial \theta ^{2}}\right\vert _{\theta =\overline{\theta }}\xi $ and $%
\left. \frac{\partial F}{\partial \theta }\right\vert _{\theta =\overline{%
\theta }}\xi $ are small when compared with $F(\overline{\theta })$ we have 
\begin{equation}
\overset{\cdot \cdot }{\xi }\ =F(\overline{\theta })=-\frac{\sin \overline{%
\theta }}{l}\sum_{i=1}^{N}A_{i}\omega _{i}^{2}\cos (\omega _{i}t)
\label{Eq:dupla_derivada_epsilon}
\end{equation}

By integrating the equation \ref{Eq:dupla_derivada_epsilon} twice with
respect to time we obtain 
\begin{equation*}
\xi (t)=\frac{\sin \overline{\theta }}{l}\sum_{i=1}^{N}A_{i}\cos (\omega
_{i}t)+c_{1}t+c_{0}
\end{equation*}

Here, $c_{0}$ and $c_{1}$ are arbitrary constants and must be assumed null
given that we want $\xi $ to be an oscillatory term, so that it should not
increase as a function of time. This is a constraint that can be imposed as
initial condition. For a set $\{\omega _{i}\}_{i=1}^{N}$ the superposition $%
\sum_{i=1}^{N}A_{i}\cos (\omega _{i}t)$ does not necessarily results in a
periodic function. For this to occur, there must be a period $T$, such that: 
$\omega _{i}T=2n_{i}\pi $ and $\omega _{j}T=2n_{j}\pi $, for every pair $%
i\neq j=1,...,N$, or simply 
\begin{equation}
\omega _{i}/\omega _{j}=n_{i}/n_{j}\text{,}  \label{Eq:Comensurable}
\end{equation}%
i.e., the ratio between frequencies must be a rational number where $n_{i}$
and $n_{j}$ are the smallest possible integers, so that, $n_{i}/n_{j}$ is an
irreducible fraction.

Let us focus our analysis only in situations where this condition is
satisfied. It is important to recall this is a problem for which an
analytical solution is not typically known. Therefore, our aim here is to
derive a stability criteria given our choices for the perturbative
functions. In order to do that, we start taking a time average of Eq. \ref%
{Eq:non-perturbated_oscilating} and assuming that the unperturbed part are
not significantly altered in a first order approximation. Within these
assumptions we find:%
\begin{equation*}
\overset{\cdot \cdot }{\overline{\theta }}\ \approx -\left. \frac{\partial U%
}{\partial \theta }\right\vert _{\theta =\overline{\theta }}+\left\langle
\left. \frac{\partial F}{\partial \theta }\right\vert _{\theta =\overline{%
\theta }}\xi \right\rangle
\end{equation*}

Where $\left\langle \cdot \right\rangle $ denotes a time average that
results in 
\begin{equation*}
\begin{array}{lll}
\left\langle \left. \frac{\partial F}{\partial \theta }\right\vert _{\theta =%
\overline{\theta }}\xi \right\rangle & = & -\frac{\sin \overline{\theta }%
\cos \overline{\theta }}{l^{2}}\left( \sum_{i=1}^{N}A_{i}^{2}\omega
_{i}^{2}\left\langle \cos ^{2}(\omega _{i}t)\right\rangle +\sum_{i\neq
j=1}^{N}A_{i}A_{j}\omega _{i}^{2}\left\langle \cos (\omega _{i}t)\cos
(\omega _{j}t)\right\rangle \right) \\ 
&  &  \\ 
& = & -\frac{\sin \overline{2\theta }}{2l^{2}}\left( \frac{1}{2}%
\sum_{i=1}^{N}A_{i}^{2}\omega _{i}^{2}(\frac{\sin \left( 2\omega
_{i}T\right) }{2T\omega _{i}}+1)+\sum_{i\neq j=1}^{N}\frac{A_{i}A_{j}\omega
_{i}^{2}}{T}\left[ \frac{\sin \left( \omega _{i}-\omega _{j}\right) T}{%
\left( \omega _{i}-\omega _{j}\right) }+\frac{\sin \left( \omega _{i}+\omega
_{j}\right) T}{\left( \omega _{i}+\omega _{j}\right) }\right] \right)%
\end{array}%
\end{equation*}

Now, given that $\overset{\cdot \cdot }{\overline{\theta }}\ =-\frac{%
\partial Ueffective}{\partial \overline{\theta }}$, where 
\begin{equation*}
U_{effective}=\frac{g}{l}\cos \overline{\theta }-\frac{\cos 2\overline{%
\theta }}{4l^{2}}\left( \frac{1}{2}\sum_{i=1}^{N}A_{i}^{2}\omega _{i}^{2}(%
\frac{\sin \left( 2\omega _{i}T\right) }{2T\omega _{i}}+1)+\sum_{i\neq
j=1}^{N}\frac{A_{i}A_{j}\omega _{i}^{2}}{T}\left[ \frac{\sin \left( \omega
_{i}-\omega _{j}\right) T}{\left( \omega _{i}-\omega _{j}\right) }+\frac{%
\sin \left( \omega _{i}+\omega _{j}\right) T}{\left( \omega _{i}+\omega
_{j}\right) }\right] \right)
\end{equation*}

the stability criteria, $\frac{\partial ^{2}}{\partial \theta ^{2}}%
U_{effective}>0$, leads to 
\begin{equation}
\left( \frac{1}{2}\sum_{i=1}^{N}A_{i}^{2}\omega _{i}^{2}(\frac{\sin \left(
2\omega _{i}T\right) }{2T\omega _{i}}+1)+\sum_{i\neq j=1}^{N}\frac{%
A_{i}A_{j}\omega _{i}^{2}}{2T}\left[ \frac{\sin \left( \omega _{i}-\omega
_{j}\right) T}{\left( \omega _{i}-\omega _{j}\right) }+\frac{\sin \left(
\omega _{i}+\omega _{j}\right) T}{\left( \omega _{i}+\omega _{j}\right) }%
\right] \right) >gl  \label{Eq:General_relation_effective_potential}
\end{equation}%
where without loss of generality, we have assumed $\overline{\theta }=0$.
Here it is important to separate our analysis in three distinct parts: $N=1$%
, $N=2$ and for an arbitrary number of cosines $N>2$.

\subsection{$N=1$;}

This case has been widely studied under different analysis \cite{Ibrahim2006}%
, \cite{Butkov2001}. When $N=1$, the result from Eq. \ref%
{Eq:General_relation_effective_potential} is simply: 
\begin{equation}
A^{2}>A_{\min }^{2}=\frac{2gl}{\omega ^{2}}  \label{Eq:. Main_relation_n=1}
\end{equation}%
valid in the small oscillations regime. As we will show in section \ref%
{Section:Results}, this stability condition is not enough to cover all of
the stability regions in a diagram $\omega \times A$. Unfortunately, it
represents only one part of the history since there is a upper bound for the
amplitude that can be observed by numerical simulations.

\subsection{$N=2$;}

This is the simplest case where the perturbative analysis cannot be
rigorously applied to all possible situations since the periodicity of the
superposition $A_{1}\cos (\omega _{1}t)+A_{2}\cos (\omega _{2}t)$ depends on
certain restrictions. As it will be shown in section \ref{Section:Results},
this case becomes particularly interesting when the amplitudes are set
equal, that is, $A_{1}=A_{2}=A$. Then, we find the condition:%
\begin{equation}
\left( \omega _{1}^{2}\frac{\sin \left( 2\omega _{1}T\right) }{2T\omega _{1}}%
+\omega _{2}^{2}\frac{\sin \left( 2\omega _{2}T\right) }{2T\omega _{2}}+%
\frac{2}{T}\left( \omega _{1}^{2}+\omega _{2}^{2}\right) \left[ \frac{\sin
\left( \omega _{1}-\omega _{2}\right) T}{\left( \omega _{1}-\omega
_{2}\right) }+\frac{\sin \left( \omega _{1}+\omega _{2}\right) T}{\left(
\omega _{1}+\omega _{2}\right) }+\frac{T}{2}\right] \right) >\frac{2gl}{A^{2}%
}\text{.}  \label{Eq. Effective_potential_N=2}
\end{equation}

If besides having equal amplitudes we also set $\omega _{1}\approx \omega
_{2}$, then we get from Eq. \ref{Eq. Effective_potential_N=2} that the
condition $A^{2}>A_{\min }^{2}/4$, certainly fulfills the requirements for
stabilization given a frequency $\omega $. However, for certain ranges of
frequencies, stabilization can be attained for amplitudes $A$ that are
slightly smaller.

\subsection{$N>2$;}

For an important and trivial case is when the frequencies are close enough $%
\omega _{1}\approx \omega _{2}\approx ...\approx \omega _{N}\approx \frac{%
2\pi }{T}$. This leads to $\sin \left( \omega _{i}-\omega _{j}\right)
T\approx \left( \omega _{i}-\omega _{j}\right) T\approx $ $0$. Moreover, we
also have $\sin 2\omega _{i}T\approx 0$ for all $i=1,...,N$. In this case we
have a simplified condition:$\ \sum_{i=1}^{N}A_{i}^{2}\omega _{i}^{2}>2gl$.
For $N$ asymptotically large and for a set of frequencies such that the $%
\{\omega _{i}\}_{i=1}^{N}$ are equally distributed according to some
probability density function $f(\omega )$ we can replace: $%
\sum_{i=1}^{N}A_{i}\cos (\omega _{i}t)\rightarrow \overline{\cos (\omega t)}%
\sum_{i=1}^{N}A_{i}$, where $\overline{\cos (\omega t)}=\int_{0}^{\infty
}\cos (\omega t)f(\omega )d\omega $ denotes the ensemble average. A standard
case, could be an uniform distribution for the frequencies choosen in an
interval $[\omega _{\min },\omega _{\max }]$, so that we can show that: 
\begin{equation*}
\overline{\cos (\omega t)}=\frac{\sin (\omega _{\max }t)-\sin (\omega _{\min
}t)}{t\ (\omega _{\max }-\omega _{\min })}
\end{equation*}

Here an interesting choice is to make $A_{i}=A(t)$, which leads to $%
\sum_{i=1}^{N}A_{i}=NA(t)$. So we have $\overline{z(t)}=NA(t)\frac{\sin
(\omega _{\max }t)-\sin (\omega _{\min }t)}{t\ (\omega _{\max }-\omega
_{\min })}$. In this case if we denote $f(t|\omega )=\frac{A(t)}{t}\sin
(\omega t)$ and so%
\begin{eqnarray*}
\frac{d^{2}f}{dt^{2}} &=&\ddot{A}(t)\sin \omega t+2\dot{A}(t)\left( \frac{%
\omega \cos \omega t}{t}-\frac{\sin \omega t}{t^{2}}\right) + \\
&&A(t)\left( \frac{2\sin \omega t}{t^{3}}-\frac{2\omega \cos \omega t}{t^{2}}%
-\frac{\omega ^{2}\sin \omega t}{t}\right) \text{.}
\end{eqnarray*}%
If $A(t)$ does not depend on time, then $f(t|\omega )\overset{t\rightarrow
\infty }{\rightarrow }0$. There is no parametric excitation and the pendulum
is asymptotically unstable as $t\rightarrow \infty $. An alternative is to
consider a linear dependence as $A(t)=Ct$. In this case, we have $\frac{%
d^{2}f}{dt^{2}}=-CN\omega ^{2}\sin \omega t+O(\frac{1}{t})$, and reescaling $%
CN=a\omega _{\max }$:%
\begin{equation*}
\overset{\cdot \cdot }{\overline{z(t)}}=\frac{-a\ \omega _{\max }^{3}}{%
\omega _{\max }-\omega _{\min }}\sin (\omega _{\max }t)+\frac{a\ \omega
_{\min }^{2}\ \omega _{\max }}{\omega _{\max }-\omega _{\min }}\sin (\omega
_{\min }t)
\end{equation*}

For the sake of the simplicity, let us consider $\omega _{\min }=0$, so that 
\begin{equation*}
\overset{\cdot \cdot }{\overline{z(t)}}=-a\omega _{\max }^{2}\sin (\omega
_{\max }t)\text{.}
\end{equation*}%
\qquad \qquad \qquad\ \ 

It is very surprising here that we recover the stability condition obtained
for the case $N=1$, by simply replacing $\omega ^{2}$ by $\omega _{\max
}^{2} $ in eq. \ref{Eq:. Main_relation_n=1}, that is:%
\begin{equation}
a^{2}>\frac{2gl}{\omega _{\max }^{2}}  \label{Eq:maximum}
\end{equation}

At this point, it is worth emphasizing that the time-dependent amplitude can
start contributing positively for stabilization by extending the length of
time in which $\cos \theta >0$. However this favorable effect soon becomes
undesirable since this monotonically increasing amplitude will dominates the
scenario leading to a loss of stability. Therefore, the real important
question to be made here is whether the system is capable of keeping the
memory of stabilization according to eq. \ref{Eq:maximum} under this
amplitude normalization procedure. We will show numerically in section \ref%
{Section:Results} that such condition is preserved. However, to show that
the condition \ref{Eq:maximum} is held, we have to analyze the survival
times instead of the survival probabilities.

\section{Numerical Simulations}

\label{Section:Numerical Simulations}

In this work, after a detailed observations in the numerical simulations, we
defined a simple stability criterion definition:

\textbf{Definition}: Given a inverted pendulum governed by eq. \ref{complete}
we say that this pendulum is stable during a window of time up to time $%
t_{\max }$ if given an initial angle $-\pi /2<\theta _{0}<\pi /2$, all $%
\theta _{t}$, $t=1,2,...t_{\max }$, obtained by integration of the motion
equations via Runge-Kutta method of fourth order satisfy:%
\begin{equation}
\cos \theta _{t}>0  \label{Eq:Main_stability_condition}
\end{equation}

This simple stability criterion leads to algorithms that although relatively
simple they can describe the stability mapping in the inverted pendulum
problem. Basically, we consider 4 procedures in our numerical simulations.
All of them are based on a main algorithm (see Table \ref%
{Table1:Main_Procedure}) which describes a generic Runge-Kutta procedure for
the inverted pendulum problem considering as input:

a) \textbf{Parametric excitation}: determined by $N$ amplitudes: $%
A[N]:(A_{1},...,A_{N})$ and $N$ frequencies $\omega \lbrack N]:(\omega
_{1},...,\omega _{N})$;

b) \textbf{Maximal number o iterations in the Rung Kutta procedure}: $%
N_{iter}$ -- This number can or cannot be attained depending on stability
condition given by eq. \ref{Eq:Main_stability_condition};

c) \textbf{Time interval for Runge Kutta iteration}: $\Delta t$

d) \textbf{Pendulum Characteristics}: $g\ $-- gravity acelaration, $l\ $--\
pendulum lengh. In this paper was considered $g=9.81$ m/s$^{2}$ and $l=1.2\ $%
m, which corresponds to a standardized broomstick lenght.

e) \textbf{Reescaling parameter}: $\nu $-- If $\nu =0$, the amplitudes are
rescaled as \ $A_{i}\rightarrow t\frac{A_{i}}{N}$, elsewhere ($\nu =1$) they
remain unchanged.

f) \textbf{External noise vector}: A string with $N_{iter}$ random Gaussian
variables with standard deviation$\ \sigma $ and mean zero.

g) \textbf{Initial conditions}: $\theta _{0}$ and $\overset{\cdot }{\theta }%
_{0}$-- Without lost of generality we consider $\overset{\cdot }{\theta }%
_{0}=0$, i.e., the pendulum starts from rest.

As \textbf{output} of this generic procedure we have:

a) \textbf{Survival time}$:$ $i$ -- Time (integer number) for which the
pendulum remains stable.

b) \textbf{Final angle}: $\theta \ $--If $\cos \theta >0$, so necessarily $%
i=N_{iter}$, elsewhere the pendulum cannot be maintained stable until the
maximal time considered as stop criteria ($N_{iter}$)

\begin{table}[tbp] \centering%
\begin{tabular}{ll}
\hline\hline
& \textbf{Main\_Runge\_Kutta\_Routine}($N,N_{iter},\Delta t,g,l,\omega
\lbrack N],A^{\ast }[N],\nu ,f[N_{iter}],\theta _{0},\dot{\theta}%
_{0},i,\theta $) \\ \hline\hline
1 & \textbf{input}:$N,N_{iter},\Delta t,g,l,\omega \lbrack N],A^{\ast
}[N],\nu ,f[N_{iter}],\theta _{0},\dot{\theta}_{0}$ \\ 
2 & \textbf{output}:$i,\theta $ \\ 
3 & \textbf{Initilizations}: $\theta =\theta _{0}$;$\ \dot{\theta}=\dot{%
\theta}_{0}$; $i=0$; \\ 
4 & \textbf{While} [($\cos (\theta )>0$).or.($i<N_{iter}$)] \textbf{do} \\ 
5 & $i:=i+1;\ t:=i\Delta t;$ \\ 
6 & $\theta _{1}:=\theta $; $\dot{\theta}_{1}=\dot{\theta}$ \\ 
7 & \ \ \ \ \ \textbf{For} $j=1,...,N$ \\ 
8 & \ \ \ \ $\ \ \ \ A_{j}=(1-v)t\ A_{j}^{\ast }/N+vA_{j}^{\ast }$ \\ 
9 & \ \ \ \ \ \textbf{Endfor} \\ 
10 & $\ \ a_{1}=\omega _{0}^{2}[1-\sum_{j=1}^{N}\frac{A_{j}\omega _{j}^{2}}{g%
}\cos (\omega _{i}t)]\sin \theta _{1}+f(i)$ \\ 
11 & $\ \ \theta _{2}:=\theta _{1}+\frac{1}{2}\dot{\theta _{1}}\Delta t$ \\ 
12 & $\dot{\ \ \theta _{2}}:=\dot{\theta _{1}}+\frac{1}{2}a_{1}\Delta t$ \\ 
13 & $\ \ a_{2}=\omega _{0}^{2}[1-\sum_{j=1}^{N}\frac{A_{j}\omega _{j}^{2}}{g%
}\cos (\omega _{i}(t+\frac{\Delta t}{2}))]\sin \theta _{2}+f(i)$ \\ 
14 & $\ \ \theta _{3}=\theta _{1}+\frac{1}{2}\dot{\theta _{2}}\Delta t$ \\ 
15 & $\dot{\ \ \theta _{3}}=\dot{\theta _{1}}+\frac{1}{2}a_{2}\Delta t$ \\ 
16 & $\ \ a_{3}=\omega _{0}^{2}[1-\sum_{j=1}^{N}\frac{A_{j}\omega _{j}^{2}}{g%
}\cos (\omega _{i}(t+\frac{\Delta t}{2}))]\sin \theta _{3}+f(i)$ \\ 
17 & $\ \ \theta _{4}=\theta _{1}+\dot{\theta _{3}}\Delta t$ \\ 
18 & $\dot{\ \ \theta _{4}}=\dot{\theta _{1}}+a_{3}\Delta t$ \\ 
19 & $\ \ a_{4}=\omega _{0}^{2}[1-\sum_{j=1}^{N}\frac{A_{j}\omega _{j}^{2}}{g%
}\cos (\omega _{i}\ast (t+\Delta t))]\sin \theta _{4}+f(i)$ \\ 
20 & $\ \ \theta =\theta +\frac{\Delta t}{6}(\dot{\theta _{1}}+2(\dot{\theta
_{2}}+\dot{\theta _{3}})+\dot{\theta _{4}})$ \\ 
21 & $\ \ \dot{\theta}=\dot{\theta}+\frac{\Delta t}{6}%
(a_{1}+2(a_{2}+a_{3})+a_{4})$ \\ 
22 & \textbf{EndWhile} \\ 
23 & \textbf{Return} $i$, $\theta $ \\ 
24 & \textbf{End} \textbf{Main\_Runge\_Kutta\_Routine} \\ \hline\hline
\end{tabular}%
\caption{Main Procedure: performs the Runge Kutta interations for the problem considering the 
all possible ingredients: excitation parameters and external noise}\label%
{Table1:Main_Procedure}%
\end{table}%

Now, we will present all procedures used in our work by reporting
specifically each one of them showing pseudo-codes.

\subsection{Procedure 1}

For $N=1$, we change $A$ and $\omega $ in the respective ranges $[A_{\min
},A_{\max }]$ and $[\omega _{\min },\omega _{\max }]$. For $\sigma =0$ we
look for each pair $(\omega ,A)$ if the pendulum stabilizes or not by
calling the main procedure (sub routine): \textbf{Main\_Runge\_Kutta\_Routine%
}. For $\sigma \neq 0$ we run $N_{run}$ times the program for different
seeds and we calculate the survival probability of pendulum, i.e., $%
p_{survival}=n_{survival}/N_{run}$, where $n_{survival}$ is the number of
times that system stabilizes. In our procedure \ref{Table:Procedure_1} $%
p_{survival}$ is denoted by $prob_{k,m}$ since it is associated to pair $%
(\omega ,A)$, parametrized as $\omega =\omega _{\min }+k\Delta \omega $ and $%
A=A_{\min }+m\Delta A$, where $k=1,...,N_{1}$ and $m=1,...,N_{2}$ (see again
the pseudo-code-- described in Table \ref{Table:Procedure_1}).

Here (and in the other procedures) $H(\theta )$ is the Heaviside function in
the cosine argument:%
\begin{equation*}
H(\theta )=\left\{ 
\begin{tabular}{lll}
1 & if & $\cos \theta >0$ \\ 
&  &  \\ 
0 & if & $\cos \theta \leq 0$%
\end{tabular}%
\right.
\end{equation*}

Similarly, $iaver_{k,m}$ corresponds to survival time average over $N_{run}$
repetitions, which is interesting only when $\sigma \neq 0$. It is important
to notice that $p_{survival}$ is either 0 or 1 when $\sigma =0$ (in this
case we make $n_{run}=1$ necessarily). Here $idum$ is the seed of uniform
random variables generator: rand$[idum]$. In this paper we used the
generator \textbf{ran2} of numerical recipes \cite{Press1992} as well as
gasdev$($rand$[idum])$ that has as input rand$[idum]$. This last routine is
the Gaussian random numbers generator according to Box-Muller method which
also is described in \cite{Press1992}.

\begin{table}[tbp] \centering%
\begin{tabular}{l}
\hline\hline
\textbf{Procedure 1 : Diagram\_}$N=1$ \\ \hline\hline
\textbf{Input:}$A_{\min },A_{\max },l$,$g$,$\omega _{\min }$,$\omega _{\max
} $,$\sigma $,$\Delta \omega $,$\Delta A,\Delta t,N_{iter},idum,N_{run}$ \\ 
\textbf{Parameters}: $N=1,\nu =1$ \\ 
$\cdot \ N_{1}=(\omega _{\max }-\omega _{\min })/\Delta \omega ;$ \\ 
$\cdot \ N_{2}=(A_{\max }-A_{\min })/\Delta A;$ \\ 
For $i_{run}=1,N_{run}$ \\ 
\ \ \ \ \ \ For $ic=1,...,N_{iter}$ \\ 
$\ \ \ \ \ \ \ \ \ \ \ \ f_{ic}\ =\ \sigma \cdot \ $gasdev$($rand$[idum])$
\\ 
\ \ \ \ \ \ EndFor \\ 
For $k=1,N_{1}$ \\ 
\ \ For $m=1,N_{2}$ \\ 
\ \ \ \ \ \ $\omega _{1}=\omega _{\min }+k\Delta \omega $ \\ 
\ $\ \ \ \ \ A_{1}^{\ast }=A_{\min }+m\Delta A$ \\ 
\textbf{Call} \textbf{Main\_Runge\_Kutta\_Routine}($N=1,N_{iter},\Delta
t,g,l,\omega \lbrack N],A^{\ast }[N],\nu ,f[N_{iter}],\theta _{0},\dot{\theta%
}_{0},i,\theta $) \\ 
\ \ \ \ \ $iaver_{k,m}=iaver_{k,m}+i/N_{run}$ \\ 
$\ \ \ \ \ \ prob_{k,m}=prob_{k,m}+H(\theta )/N_{run}$ \\ 
\ \ EndFor \\ 
EndFor \\ 
EndFor \\ 
For $k=1,N_{1}$ \\ 
\ \ For $m=1,N_{2}$ \\ 
\ \ \ \ \ \ $freq=\omega _{\min }+k\Delta \omega $ \\ 
\ $\ \ \ \ Ampl=A_{\min }+m\Delta A$ \\ 
\ \ \ \ \ Print $freq,Ampl,iaver_{k,m},prob_{k,m}$ \\ 
\ \ EndFor \\ 
EndFor \\ 
\textbf{End Procedure} \\ \hline\hline
\end{tabular}%
\caption{This procedure is used to build data for a diagram of survival probability for each pair 
($\omega$, $A$) considering the parametric excitation (oscillation at the suspention) with um cossine ($N=1$)
and an additive (white) noise}\label{Table:Procedure_1}%
\end{table}%

\subsection{Procedure 2}

For $N=2$, we fix $A_{1}=A_{2}=A\ $and we pick up $\omega _{1}$ and $\omega
_{2}$ by chance. When $\sigma =0$, for each pair $(\omega _{1},\omega _{2})$
spanned in the invervals $[\omega _{\min }^{(1)},\omega _{\max }^{(1)}]$ and 
$[\omega _{\min }^{(2)},\omega _{\max }^{(2)}]$ respectively, we look
whether the pendulum stabilizes or not. For $\sigma \neq 0$ we run $N_{run}$
times the program for different seeds and we estimate the pendulum survival
probability, i.e., $p_{survival}=n_{survival}/N_{run}$, as shown in
procedure 1. This procedure can be observed in pseudo-code described in
table \ref{Table:Procedure_2}.

\begin{table}[tbp] \centering%
\begin{tabular}{l}
\hline\hline
\textbf{Procedure 2: Diagram\_}$N=2$ \\ \hline\hline
\textbf{Input:}$A$,$l$,$g$,$\omega _{\min }^{(1)}$,$\omega _{\max }^{(1)}$,$%
\omega _{\min }^{(2)}$,$\omega _{\min }^{(2)}$,$\sigma $,$\Delta \omega $,$%
\Delta t,N_{iter},idum,N_{run}$ \\ 
\textbf{Parameters}: $N=2$, $A_{1}^{\ast }=A$;$\ A_{2}^{\ast }=A;\nu =1$ \\ 
$\cdot \ N_{1}=(\omega _{\max }^{(1)}-\omega _{\min }^{(1)})/\Delta \omega ;$
\\ 
$\cdot \ N_{2}=(\omega _{\max }^{(2)}-\omega _{\min }^{(2)})/\Delta \omega ;$
\\ 
For $i_{run}=1,N_{run}$ \\ 
\ \ \ \ \ \ For $ic=1,...,N_{iter}$ \\ 
$\ \ \ \ \ \ \ \ \ \ \ \ f_{ic}=\sigma \cdot $gasdev$($rand$[idum])$ \\ 
\ \ \ \ \ \ EndFor \\ 
For $k=1,N_{1}$ \\ 
\ \ For $m=1,N_{2}$ \\ 
\ \ \ \ \ \ $\omega _{1}=\omega _{\min }^{(1)}+k\Delta \omega $ \\ 
\ $\ \ \ \ \ \omega _{2}=\omega _{\min }^{(2)}+m\Delta \omega $ \\ 
\textbf{Call} Main\_Sub\_Routine($N,N_{iter},\Delta t,g,l,\omega \lbrack
N],A^{\ast }[N],\nu ,f[N_{iter}],\theta _{0},\dot{\theta}_{0},i,\theta $) \\ 
\ \ \ \ \ $iaver_{k,m}=iaver_{k,m}+i/N_{run}$ \\ 
$\ \ \ \ \ \ prob_{k,m}=prob_{k,m}+H(\theta )/N_{run}$ \\ 
\ \ EndFor \\ 
EndFor \\ 
EndFor \\ 
For $k=1,N_{1}$ \\ 
\ \ For $m=1,N_{2}$ \\ 
\ \ \ \ \ \ $freq1=\omega _{\min }^{(1)}+k\Delta \omega $ \\ 
\ $\ \ \ \ \ freq2=\omega _{\min }^{(2)}+m\Delta \omega $ \\ 
\ \ \ \ \ Print $freq1,freq2,iaver_{k,m},prob_{k,m}$ \\ 
\ \ EndFor \\ 
EndFor \\ 
\textbf{End\_Procedure} \\ \hline\hline
\end{tabular}%
\caption{This procedure produces data for the diagram of survival probability for each pair 
($\omega_{1}$, $\omega_{2}$) considering the parametric excitation (oscillation at the suspention) with
a superposition of two  cosines ($N=2$) and an additive (white) noise. Here the amplitudes are $A_{1}=A_{2}=A$, 
which is also a input of the algorithm}\label{Table:Procedure_2}%
\end{table}%

\subsection{Procedure 3}

Here we analyze the problem with arbitrary $N>2$. More precisely, we analyze
the effects for $N\rightarrow \infty $ by a amplitude renormalization such
that $A(t)\rightarrow \frac{At}{N}$. It is worth notice that in doing so, we
realize this is a similar problem to $N=1$, as previously described in
section \ref{Section:Perturbative_Analysis}. For each selected $A$ which
varies in the range $[A_{\min },A_{\max }]$ according to lag $\Delta A$, we
attribute $A_{1}=A_{2}=...=A_{N}=A$ and we choosen $N$ random uniform
variables $\omega _{1}$, $\omega _{2}$, ...., $\omega _{N}$ in the inverval [%
$0$,$\omega _{\max }$]. The value maximum frequency to be drawn $\omega
_{\max }$, assumes values in the interval [$0,\omega _{\max }^{\sup }$]
varying according to a shift $\Delta \omega $. So this procedure calls the
main sub-routine Table: \ref{Table1:Main_Procedure} with $\nu =0$ (which
makes the rescaling). Here $N$ is an arbitrary input, since we study the
effects of $N$ in the asymptotic limit $N\rightarrow \infty $. \ 

In this case, the algorithm with this rescaling, computes the survival time
(the time up destabilization) in order to compare with stabilization
diagrams with $N=1$. This procedure is summarized according to pseudo-code
described in table \ref{Table:procedure_3}.

\begin{table}[tbp] \centering%
\begin{tabular}{l}
\hline\hline
\textbf{Procedure 3 : Reescaling} \\ \hline\hline
\textbf{Input:}$A_{\min },A_{\max }$,$N$,$l$,$g$,$\omega _{\max }^{(\sup )}$,%
$\Delta \omega $,$\Delta A,\Delta t,N_{iter},idum,N_{run}$ \\ 
\textbf{Parameters}: $\nu =0$ \\ 
$\cdot \ N_{2}=(A_{\max }-A_{\min })/\Delta A;$ \\ 
$\cdot \ N_{1}=\omega _{\max }^{(\sup )}/\Delta \omega $ \\ 
For $i_{run}=1,N_{run}$ \\ 
\ For $k=1,N_{1}$ \\ 
$\ \ \ \ \ \ \omega _{\max }=k\Delta \omega $ \\ 
\ \ \ \ \ \ \ \ \ \ \ For $i=1,N$ \\ 
$\ \ \ \ \ \ \ \ \ \ \ \ \ \ \ \omega _{i}=$rand$[idum]\cdot \omega _{\max }$
\\ 
\ \ \ \ \ \ \ \ \ \ \ End For \\ 
\ \ \ For $m=1,N_{2}$ \\ 
\ $\ \ \ \ \ Aux=A_{\min }+m\Delta A$ \\ 
\ \ \ \ \ \ \ \ \ \ For $i=1,N$ \\ 
\ \ \ \ \ \ \ \ \ \ \ \ \ \ \ $A_{i}^{\ast }=Aux$ \\ 
\ \ \ \ \ \ \ \ \ \ EndFor\  \\ 
\ \ \ \ \ \ \ \ \ \textbf{Call} Main\_Sub\_Routine($N,N_{iter},\Delta
t,g,l,\omega \lbrack N],A^{\ast }[N],\nu ,f[N_{iter}],\theta _{0},\dot{\theta%
}_{0},i,\theta $) \\ 
\ \ \ \ \ $iaver_{k,m}=iaver_{k,m}+i/N_{run}$ \\ 
$\ \ \ \ \ \ prob_{k,m}=prob_{k,m}+H(\theta )/N_{run}$ \\ 
\ \ \ \ EndFor \\ 
\ EndFor \\ 
EndFor \\ 
For $k=1,N_{1}$ \\ 
\ \ For $m=1,N_{2}$ \\ 
\ \ \ \ \ \ $freq=k\Delta \omega $ \\ 
\ $\ \ \ \ Ampl=A_{\min }+m\Delta A$ \\ 
\ \ \ \ \ Print $freq,Ampl,iaver_{k,m},prob_{k,m}$ \\ 
\ \ EndFor \\ 
EndFor \\ 
\textbf{End Procedure} \\ \hline\hline
\end{tabular}%
\caption{Giving a superposition of $N$ cosines exciting the basis of pendulum, this procedure 
calculate the average survival time of pendulum calling the main subroutine when the amplitudes are
rescaled. The plots must recover in some instance, the plots the standard plots for $N=1$}%
\label{Table:procedure_3}%
\end{table}%

\subsection{Procedure 4}

Finally, we look for an optimum number of cosines $N$ in the stabilization
of the inverted pendulum. For arbitrary $N$, we also perform an optimization
algorithm. Given the frequencies $\omega _{1}$, $\omega _{2}$, ..., $\omega
_{N}$ and $A_{1}$, $A_{2}$, ...., $A_{N}$ randomly chosen uniformily in the
respective intervals $[\omega _{\min },\omega _{\max }]$ and $[A_{\min
},A_{\max }]$, we search for the number $N$ that maximizes the stabilization
probability. Therefore the algorithm run $N_{run}$ different formulas with
parametric excitation $z(t)=\sum_{i=1}^{N}A_{i}\cos (\omega _{i}t)$ and call
the main sub-routine that solves the Runge-Kutta for each set: $%
\{(A_{1},\omega _{1}),...,(A_{N},\omega _{N})\}$. From that, we calculate
the $p_{survival}=n_{survival}/N_{run}$. The procedure also computes the
average time survival for completeness, but it is not used in this work.

\begin{table}[tbp] \centering%
\begin{tabular}{l}
\hline\hline
\textbf{Procedure 4 : Optimization} \\ \hline\hline
\textbf{Input:}$A_{\min },A_{\max }$,$N_{\max }$,$l$,$g$,$\omega _{\min }$,$%
\omega _{\max }$,$\Delta \omega $,$\Delta A,\Delta
t,N_{iter},idum1,idum2,N_{run}$ \\ 
\textbf{Parameters}: $\nu =1$ \\ 
For $N=1,N_{\max }$ \\ 
\ \ \ For $i_{run}=1,N_{run}$ \\ 
\ \ \ \ \ \ \ \ \ \ \ For $i=1,N$ \\ 
$\ \ \ \ \ \ \ \ \ \ \ \ \ \ \ \omega _{i}=\omega _{\min }+$rand$%
[idum1]\cdot (\omega _{\max }-\omega _{\min })$ \\ 
$\ \ \ \ \ \ \ \ \ \ \ \ \ \ A_{i}=A_{\min }+$rand$[idum2]\cdot (A_{\max
}-A_{\min })$ \\ 
\ \ \ \ \ \ \ \ \ \ \ End For \\ 
\ \ \ \ \ \ \ \ \ \textbf{Call} Main\_Sub\_Routine($N,N_{iter},\Delta
t,g,l,\omega \lbrack N],A^{\ast }[N],\nu ,f[N_{iter}],\theta _{0},\dot{\theta%
}_{0},i,\theta $) \\ 
\ \ \ \ \ $\ \ \ \ \ \ iaver_{N}=iaver_{N}+i/N_{run}$ \\ 
$\ \ \ \ \ \ \ \ \ \ \ prob_{N}=prob_{N}+H(\theta )/N_{run}$ \\ 
\ \ \ \ EndFor \\ 
EndFor \\ 
For $N=1,N_{\max }$ \\ 
\ \ \ \ \ Print $N,iaver_{N},prob_{N}$ \\ 
EndFor \\ 
\textbf{End Procedure} \\ \hline\hline
\end{tabular}%
\caption{Procedure that determines the $N$ that maximizes the probability of stabilization considering different
ensemble of formulas}\label{Table:Procedure_4}%
\end{table}%

\section{Results}

\label{Section:Results}

First of all we start looking at the phase diagrams for $N=1,$ $\frac{%
d^{2}\theta }{dt^{2}}=\frac{g}{l}\left( 1-\frac{1}{g}A\omega ^{2}\cos
(\omega t)\right) \sin \theta +\xi (t)$, in Fig. \ref%
{Fig:time_evolving_no_convergence}. Here we will show that our simple
stability criteria $\cos \theta (t)>0\ $is in accordance with results
obtained from literature (see for example \cite{Ibrahim2006}\cite{Butkov2001}%
)) which are based on perturbative analysis as shown in section \ref%
{Section:Perturbative_Analysis}. Based on our stability criterium we
initially integrate the equations according to algorithms described in
section \ref{Section:Numerical Simulations}, in order to check the main
results and to verify some important points not explored in literature yet.
The results for $N=1$ are also important to give insights to the other cases
($N\geq 2$).

\begin{figure}[th]
\begin{center}
\includegraphics[width=\columnwidth]{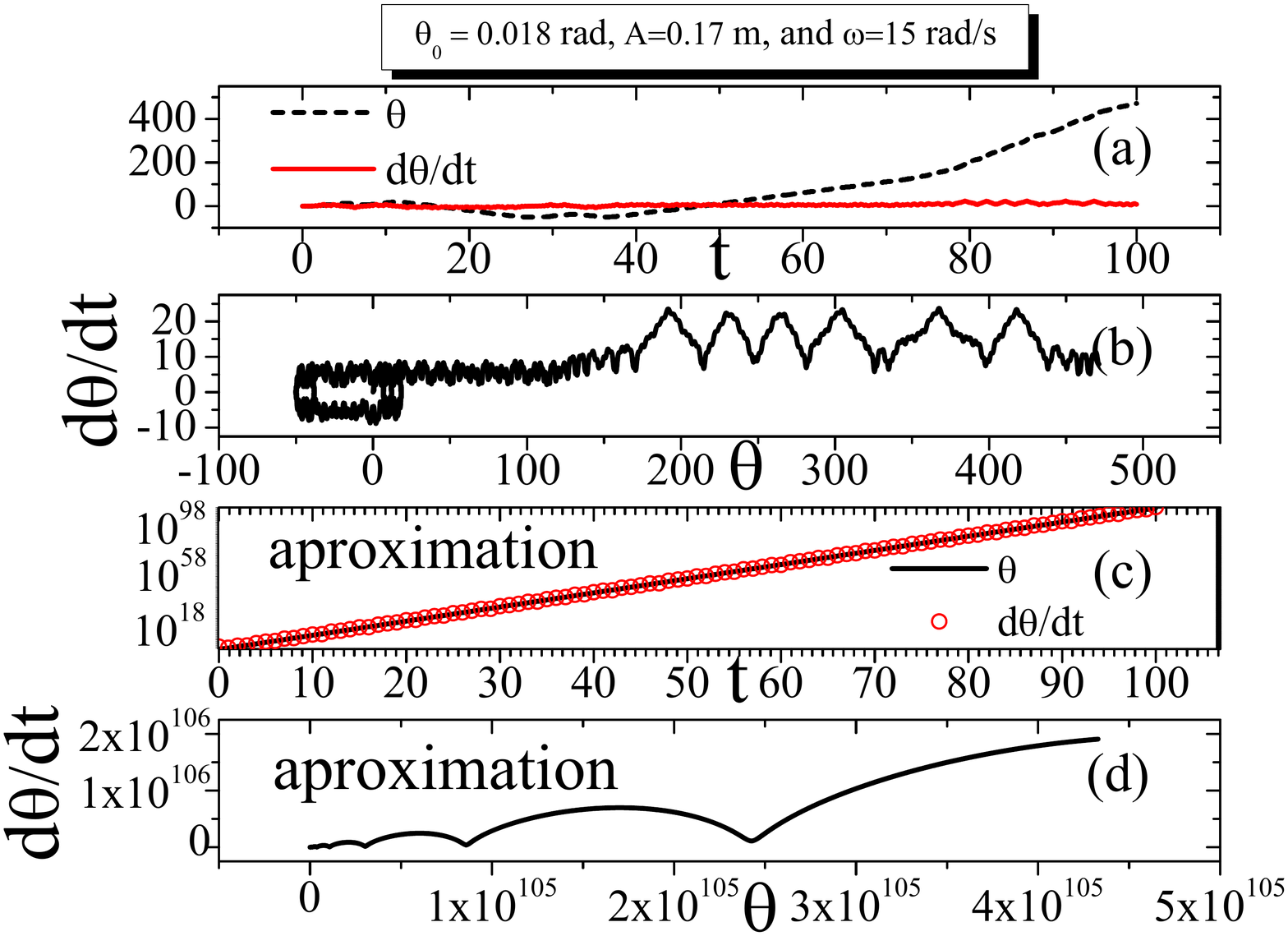}
\end{center}
\caption{Results for time evolving and phase diagram considering $\protect%
\theta _{0}=0.018$ rad, $\protect\omega =15$ rad/s and $A=0.17$ m. (a) Time
evolving $\protect\theta $ and $\protect\overset{\cdot }{\protect\theta }$.
(b) Corresponding phase diagram $\protect\overset{\cdot }{\protect\theta }%
\times \ \protect\theta $ (c) The corresponding Fig. (a) in small
oscillations approximation $\sin \protect\theta \approx \protect\theta $--
Mathieu equation. (d) Corresponding phase diagram in this approximation. }
\label{Fig:time_evolving_no_convergence}
\end{figure}

In Fig. \ref{Fig:time_evolving_no_convergence} we show results of
simulations starting from a small angle, $\theta _{0}=0.018$, and using
frequency $\omega =15$ rad/s and $A=0.17$ m. In our simulations, $l=1.2$ m
and $g=9.81$ m/s$^{2}$which brings in our imagination the typical situation
of a child trying to stabilize a broomstick on its hand. Other dimensions
deserve discussion for large $N$ which will be considered in other
contribution \cite{SilvaPrado2016}.

In all the following results, we have used $t_{\max }=10^{6}$ iterations and
\ $\varepsilon =\Delta t=10^{-5}$. These parameter values were settled after
the observation that for $t\geq t_{\max }$ and $\varepsilon ^{\prime
}<\varepsilon $ no significant variations were detected. Fig. (\ref%
{Fig:time_evolving_no_convergence} a) and (\ref%
{Fig:time_evolving_no_convergence} b) show the time evolution and the
corresponding phase diagram respectively. In this simulation $\xi (t)=0$, so
that there is not any stochastic noise. The corresponding plots to (\ref%
{Fig:time_evolving_no_convergence} a) and (\ref%
{Fig:time_evolving_no_convergence} b) when we use the small angle
approximation $\sin \theta \approx \theta $ (known as Mathieu equation \cite%
{Mathieu1947})

\begin{equation}
\overset{\cdot \cdot }{\theta }-\frac{g}{l}\left( 1-\frac{1}{g}A\omega
^{2}\cos (\omega t)\right) \theta =0  \label{Eq:Mathieu_equation}
\end{equation}%
are observed in (\ref{Fig:time_evolving_no_convergence} c) and (\ref%
{Fig:time_evolving_no_convergence} d) respectively.

The fact that this figure illustrates a case where the initial condition
leads to a non-stable outcome is not os relevant here, since they do not
satisfy eq. \ref{Eq:. Main_relation_n=1}. However what call ones attention
is the fact that small initial angle leads to a different different
divergence for $\theta (t)$ for large times (instability) whether the one
replaces $\sin \theta $ by $\theta $ or not. The plot (\ref%
{Fig:time_evolving_no_convergence} c) is in mono-log scale since there is a
exponential divergence (straight line in this scale) which is more
pronounced than in (\ref{Fig:time_evolving_no_convergence} a). Such aspect
although seems very simple is simply discarded by some authors in
literature. The very different phase diagrams (b) and (d) obtained for these
different regimes shows even more our thesis about this topic.

\begin{figure}[th]
\begin{center}
\includegraphics[width=\columnwidth]{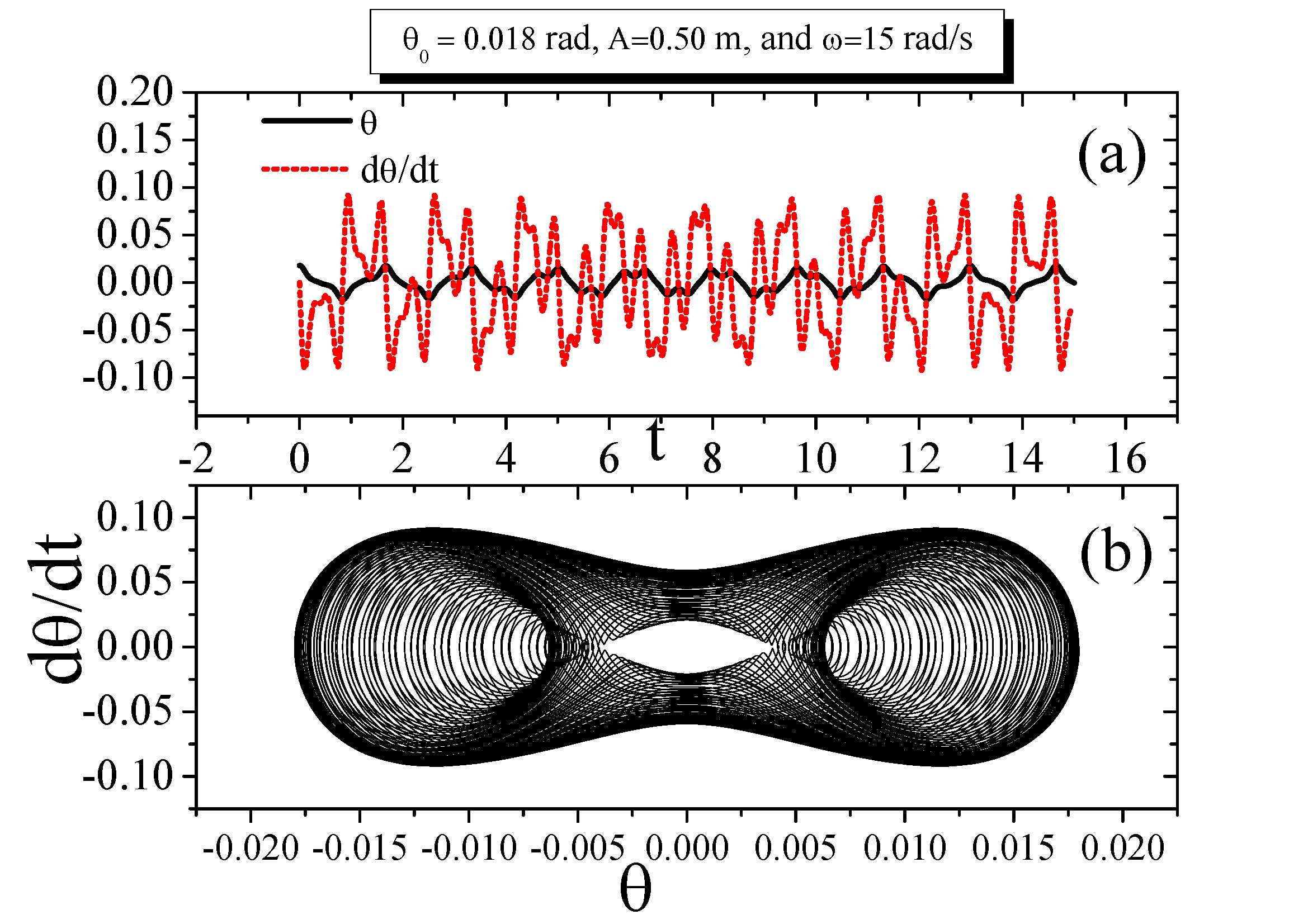}
\end{center}
\caption{Results for time evolving and phase diagram considering $\protect%
\theta _{0}=0.018$ rad, $\protect\omega =15$ rad/s and $A=0.50$ m. (a) Time
evolving $\protect\theta $ and $\protect\overset{\cdot }{\protect\theta }$.
(b) Corresponding phase diagram $\protect\overset{\cdot }{\protect\theta }%
\times \ \protect\theta $. }
\label{Fig:time_evolving_convergence}
\end{figure}

In the results shown in Fig. \ref{Fig:time_evolving_no_convergence}, we
consider a larger amplitude $A=0.50$ m. Now, that Eq. \ref{Eq:.
Main_relation_n=1} is satisfied, we can see a periodic behavior for $\theta $
and $\overset{\cdot }{\theta }$ as function of time in Fig. (\ref%
{Fig:time_evolving_convergence} a) and now beautiful Lissajous plot in the
phase space shown in Fig. (\ref{Fig:time_evolving_convergence} b).

Differently from the Fig. \ref{Fig:time_evolving_no_convergence} the plots (%
\ref{Fig:time_evolving_convergence}c) and (\ref%
{Fig:time_evolving_convergence}d) corresponding to small oscillations regime
are here omitted since there is no significant difference in the simulations.

\begin{figure}[th]
\begin{center}
\includegraphics[width=\columnwidth]{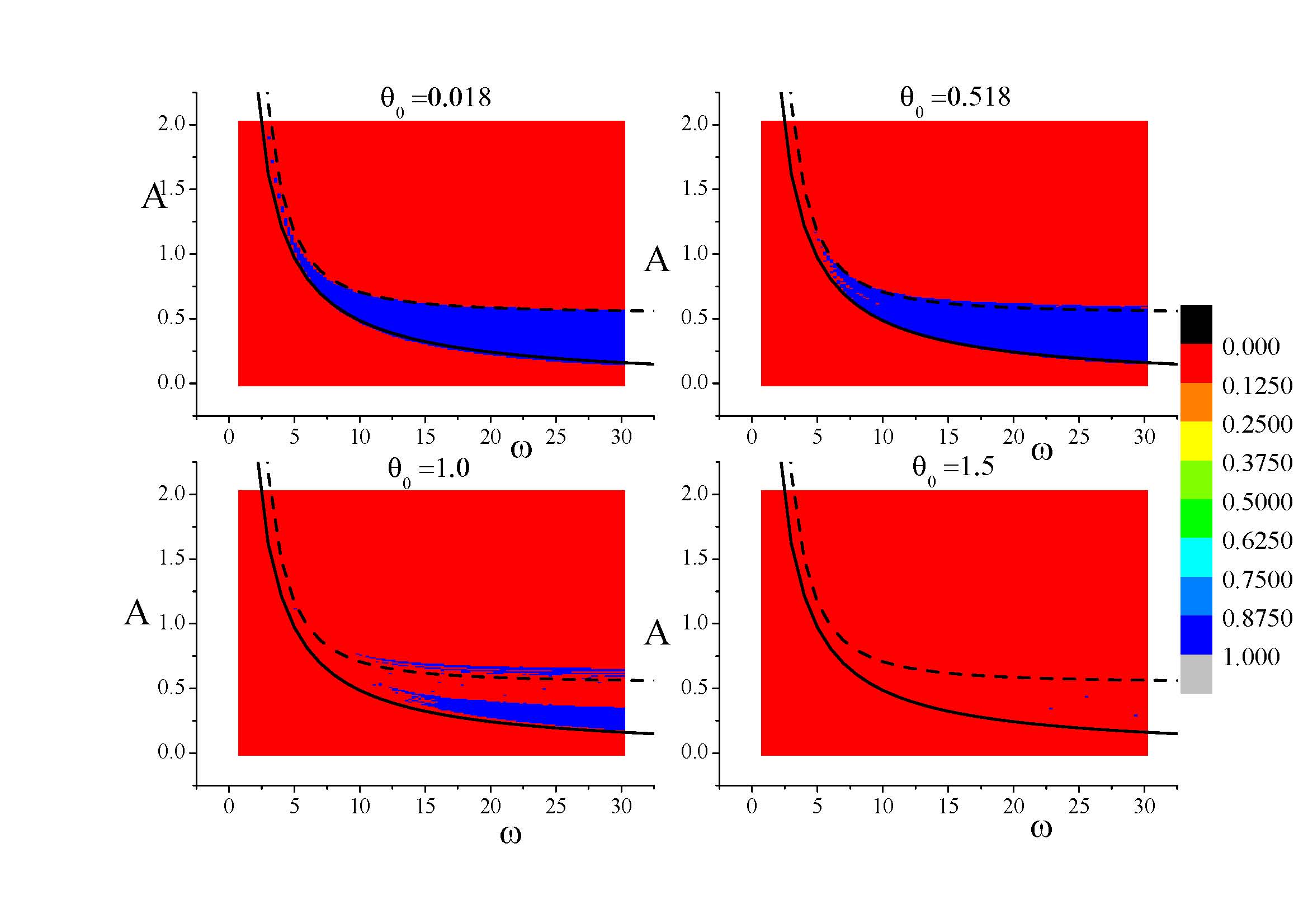}
\end{center}
\caption{Initial angle effects on the diagram phases $\protect\omega \times
A $. The upper (solid) and lower (dashed) curves correspond respectively to
the limits established by Eqs. \protect\ref{Eq:. Main_relation_n=1} and 
\protect\ref{Eq.parametric_ressonance}. }
\label{Fig:Initial_angle_effects_one_cossine}
\end{figure}

In order to study initial angles' effects, we analyze the phase diagram $%
\omega \times A$ obtained via numerical simulations. The diagrams are shown
in Fig. \ref{Fig:Initial_angle_effects_one_cossine}, and which they
correspond to results from Procedure 1: Table \ref{Table:Procedure_1} (in
this case we make $N_{run}=1$). It is important to consider that Eq. \ref%
{Eq:. Main_relation_n=1} determines a lower bound for amplitude: $A_{\min }=%
\frac{\sqrt{2gl}}{\omega }$. On the other hand, when the amplitude $A$ is
increased beyond a certain critical value $A_{\max }$, the pendulum loses
its stability again \cite{Blackburn1992,Smith1992,Butkov2001} and its
evolution cannot be described by effective potential method (perturbative
analysis). It is shown in Butikov \cite{Butkov2001}, based in simulation
(heuristic) arguments, shows that the solution over the upper boundary of
stability has a simple spectral decomposition in only two frequencies: $%
\omega /2$ and $3\omega /2$, such that $\theta (t)=A_{1}\cos (\omega
t/2)+A_{3}\cos (3\omega t/2)$. By using this hypotesis and substituting this
solution in Eq. \ref{Eq:Mathieu_equation} (instead of the exact equation Eq. %
\ref{complete}) we have: 
\begin{equation}
A<A_{\max }=\frac{l}{4}\left[ \sqrt{117+232(\omega _{0}/\omega
)^{2}+80(\omega _{0}/\omega )^{4}}-9-4(\omega _{0}/\omega )^{2}\right]
\label{Eq.parametric_ressonance}
\end{equation}

The solid and dashed black curves show respectively the lower (Eq. \ref{Eq:.
Main_relation_n=1}) and upper (Eq. \ref{Eq.parametric_ressonance}) stability
domain boundaries. First, we see that the blue region is being destroyed as
the initial angle increases, but we need to pay attention to the way it
happens. We can observe an interesting effect: there is a set of conditions
in the primary stability region that loses its stability so that the region
becomes fragmented, while another set above the upper limit becomes stable.
The upper limit, established in \cite{Butkov2001}, is really restricted to
small angles showing that is based in the approximated equation, which again
indicates the importance of the numerical work here. However the lower bound
obtained by the effective potential is absolutely respected\ (not invaded by
stability region).

\begin{figure}[th]
\begin{center}
\includegraphics[width=\columnwidth]{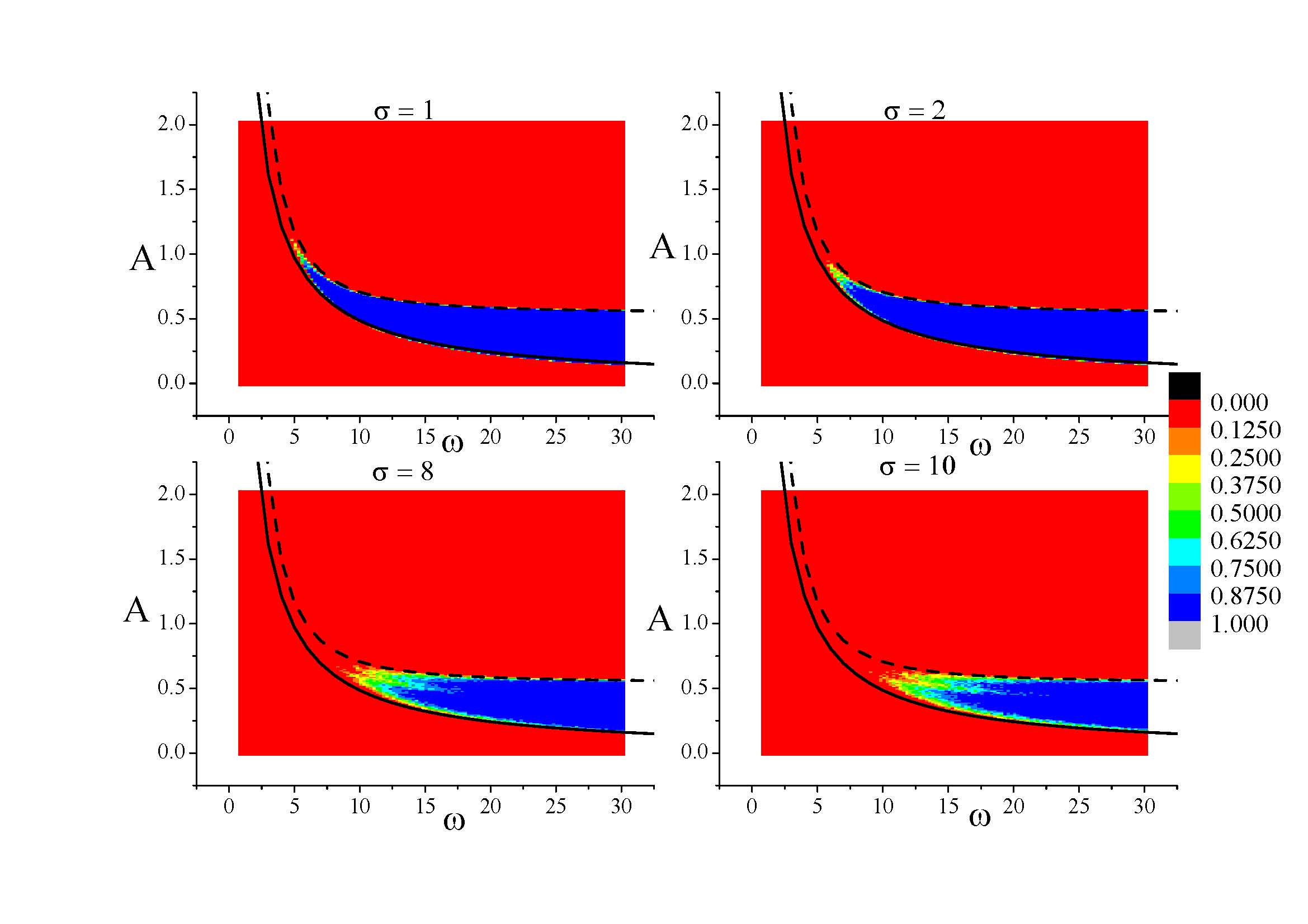}
\end{center}
\caption{External (additive) stochastic noise effects on the diagram phases $%
\protect\omega \times A$. The different plots correspond to different
standard deviations ($\protect\sigma $) of noise $\protect\phi (t)$.}
\label{Fig:noise_effects_one_cossine}
\end{figure}

In Fig. \ref{Fig:noise_effects_one_cossine} we show the effects of an
additive random noise. We are interested in seeing how the stability diagram
is degraded according to the increase of noise variance. Each plot
corresponds to a different variance ($\sigma ^{2}$). So, we run procedure I
(see table \ref{Table:Procedure_1}) with $N_{run}\ $= 50 times with
different seeds and we calculate the survival probability:%
\begin{equation*}
p_{survival}=\frac{n_{survival}}{N_{run}}
\end{equation*}%
where $n_{survival}$ is the number of times in which our pendulum
stabilizes. The color scale are graduated according to the $p_{survival}$%
-values that were obtained.

Now we focus our analysis in the case $N=2$, where we make $A_{1}=A_{2}=A$
according to Procedure II: Table \ref{Table:Procedure_2}. The results show a
rich structure as we can see in Fig \ref{Fig:Neq2_A=0,17_diff_initial_angles}%
. We illustrate two different initial angles for $A=0.17$ recalling that red
stands for unstable regions, while blue denotes stabilization. Comparing the
figures on the left with the ones on the right (small angles approximation)
we can see that it is important to consider $\sin \theta $ and not to make
the approximation $\sin \theta \approx \theta $, even for very small angles $%
\theta _{0}=0.018$ rad $\approx 1^{o}$. We can observe a less restrictive
condition $\omega _{1}^{2}+\omega _{2}^{2}\leq \frac{2gl}{A^{2}}$ (a quarter
circle, plotted in all figures), which corresponds a condition that $\omega
_{1}\approx \omega _{2}=\omega $. Exactly in diagonal the condition goes to $%
\omega ^{2}\leq \frac{gl}{A^{2}}$, which asserts the diagonal line
penetrating the 1/4-circle. But, we have more stability regions inside this
semi-circle which depend on the proximity of the diagonal. However for $%
\theta _{0}=0.518$ the system recover the restriction and all quarter of
circle is completed but not for the small angles approximation.

However, even more interesting, one should note that it is not always the
case that $\omega _{1}$ and $\omega _{2}$, both large, will lead to
stabilization. We see branches of unstable regions that remind us of Arnold
tongues \cite{Arnold1989} breaking the stability sea, specially around
diagonal the diagonal. It is not our task in this paper to describe the
properties of these unstable branches, but they are certainly very rich sets
of fractal dimensions \cite{Ott1993}.

This fractal structure set is deeply modified when $\theta _{0}=0.518$ but
only when the numerical solution is not performed in small angle
approximations.

\begin{figure}[th]
\begin{center}
\includegraphics[width=\columnwidth]{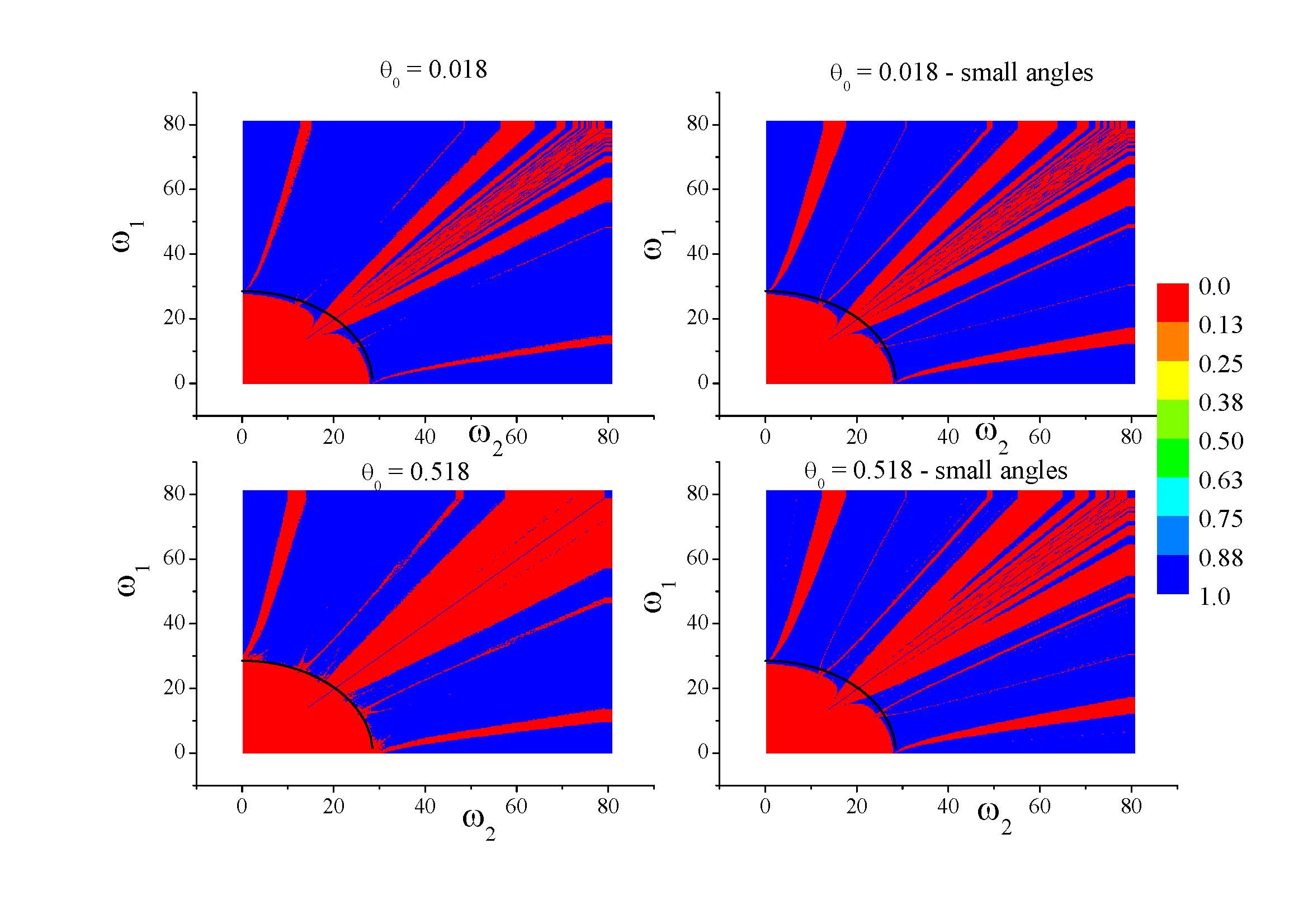}
\end{center}
\caption{Initiall angle effects $N=2$. We used $A=0.17.$}
\label{Fig:Neq2_A=0,17_diff_initial_angles}
\end{figure}
In the Fig. \ref{Fig:Neq2_A=0,34_diff_initial_angles} we show the same
simulation of Fig. \ref{Fig:Neq2_A=0,17_diff_initial_angles} for a larger
amplitude $A=0.34$. Now we have a larger instability region. Details of the
complexity of unstable branches look less pronounced on the scale of this
figure, but it does not mean that they are not there. 
\begin{figure}[th]
\begin{center}
\includegraphics[width=\columnwidth]{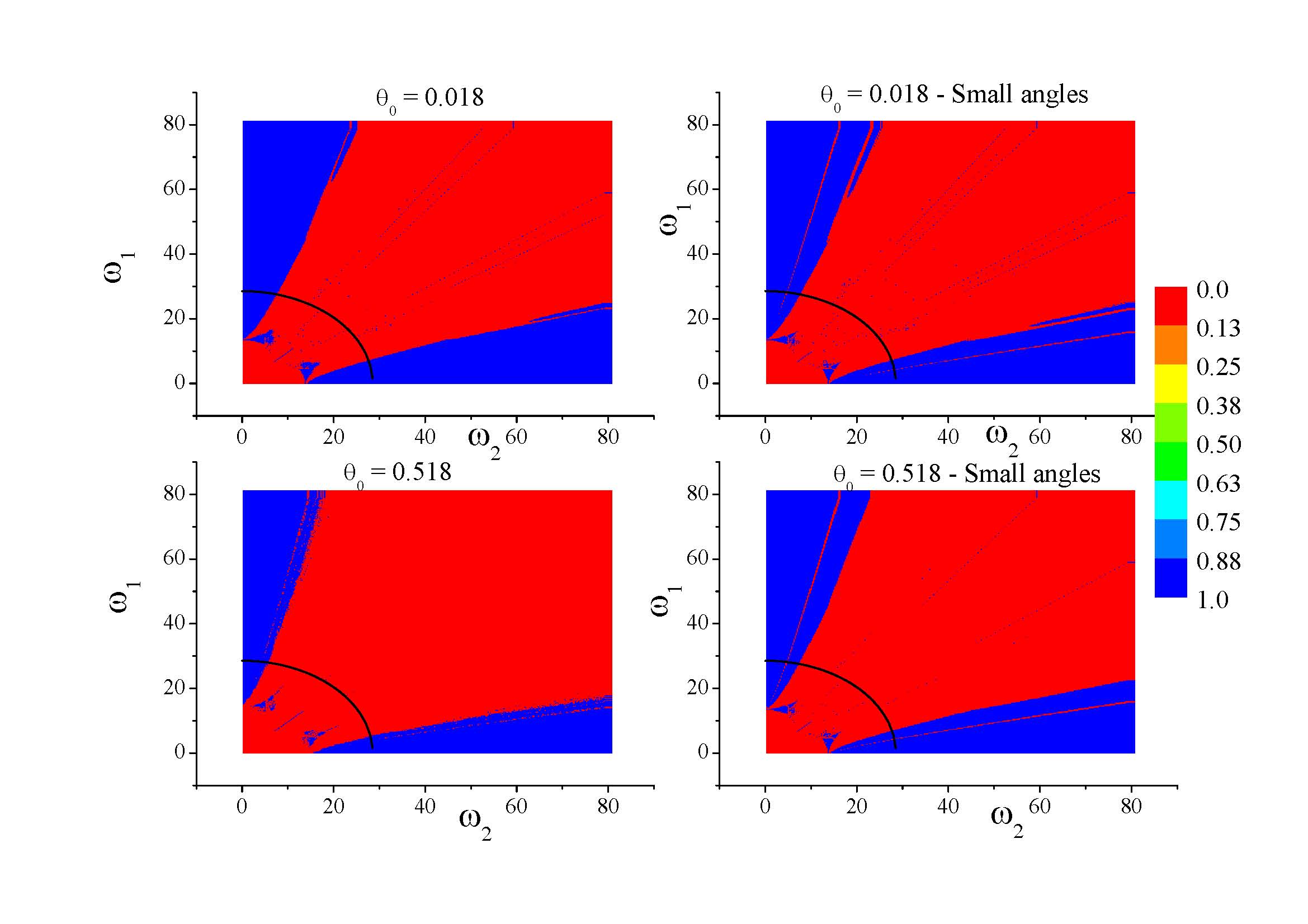}
\end{center}
\caption{Initiall angle effects $N=2$. We used $A=0.34$. Similar to Fig. 
\protect\ref{Fig:Neq2_A=0,17_diff_initial_angles}.}
\label{Fig:Neq2_A=0,34_diff_initial_angles}
\end{figure}

\begin{figure}[th]
\begin{center}
\includegraphics[width=\columnwidth]{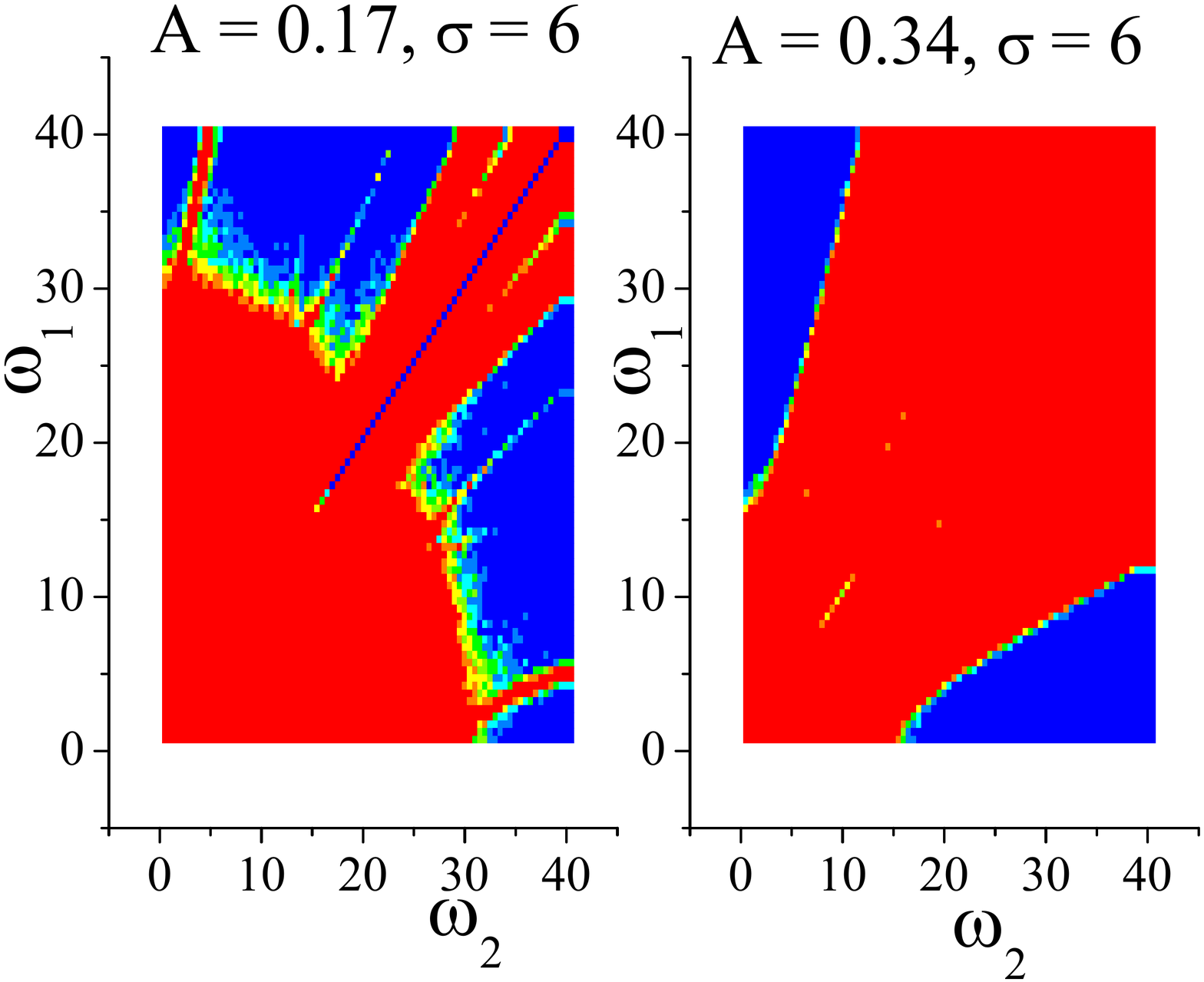}
\end{center}
\caption{External Noise effects for effects $N=2$. Comparison for $A=0.17$
and $A=0.34$ for $\protect\sigma =6$. We consider $\protect\theta _{0}=0.018$%
. }
\label{Fig:Degradation_N=2}
\end{figure}

In Fig. \ref{Fig:Degradation_N=2} we also analyze the effect of noise on the
stabilization diagram similar to the case shown in Fig. \ref%
{Fig:noise_effects_one_cossine} for $N=1$. Again, $N_{run}=50$ and we
observe only one case one case ($\sigma =6$), since the behavior is similar
to the case $N=1$, that is, the degraded region enlarges as $\sigma $
enlarges. The case $A=0.34\ $m is less sensitive to degradation than $A=0.17$
m.

\begin{figure}[th]
\begin{center}
\includegraphics[width=\columnwidth]{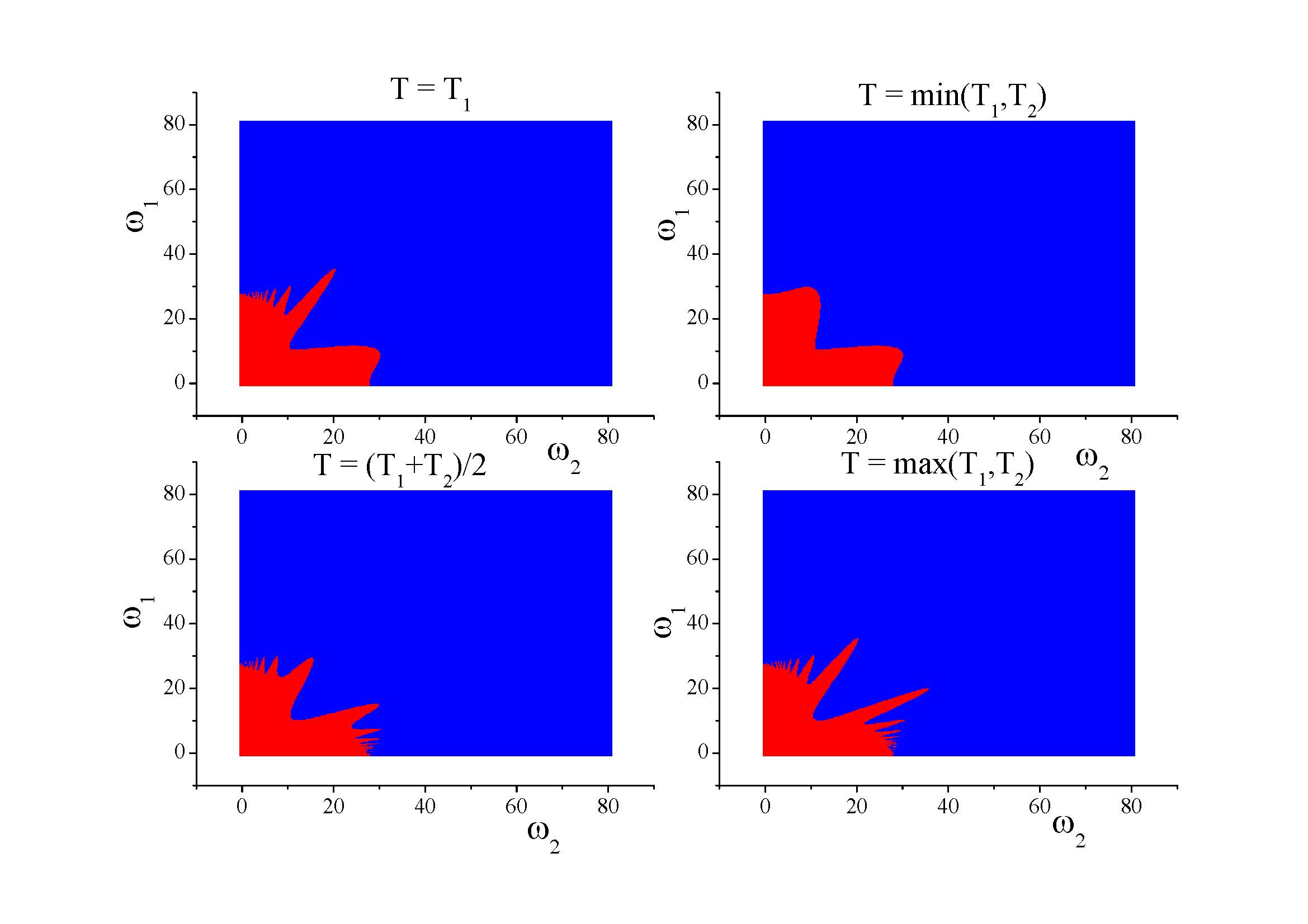}
\end{center}
\caption{Potential effective for $N=2$ according to Eq. \protect\ref{Eq.
Effective_potential_N=2}. We show some choices for $T$. The best one (that
better fits with simulations is to consider the maximal between periods. }
\label{Fig: Effective_best_T}
\end{figure}

At this point it is important to analyse the results from the perspective of
the effective potential method according to Eq. \ref{Eq.
Effective_potential_N=2}. \ Recall that we are dealing with the case of $N=2$
and equal amplitudes so that Eq. \ref{Eq. Effective_potential_N=2} depends
on a variable $T$, which must be function of $\omega _{1}$ and $\omega _{2}$
not always easily determined. So, first we consider what we think that are
reasonable choices of $T$ as shown in Fig. \ref{Fig: Effective_best_T}.  
\qquad \qquad\ \ 

\begin{figure}[th]
\begin{center}
\includegraphics[width=%
\columnwidth]{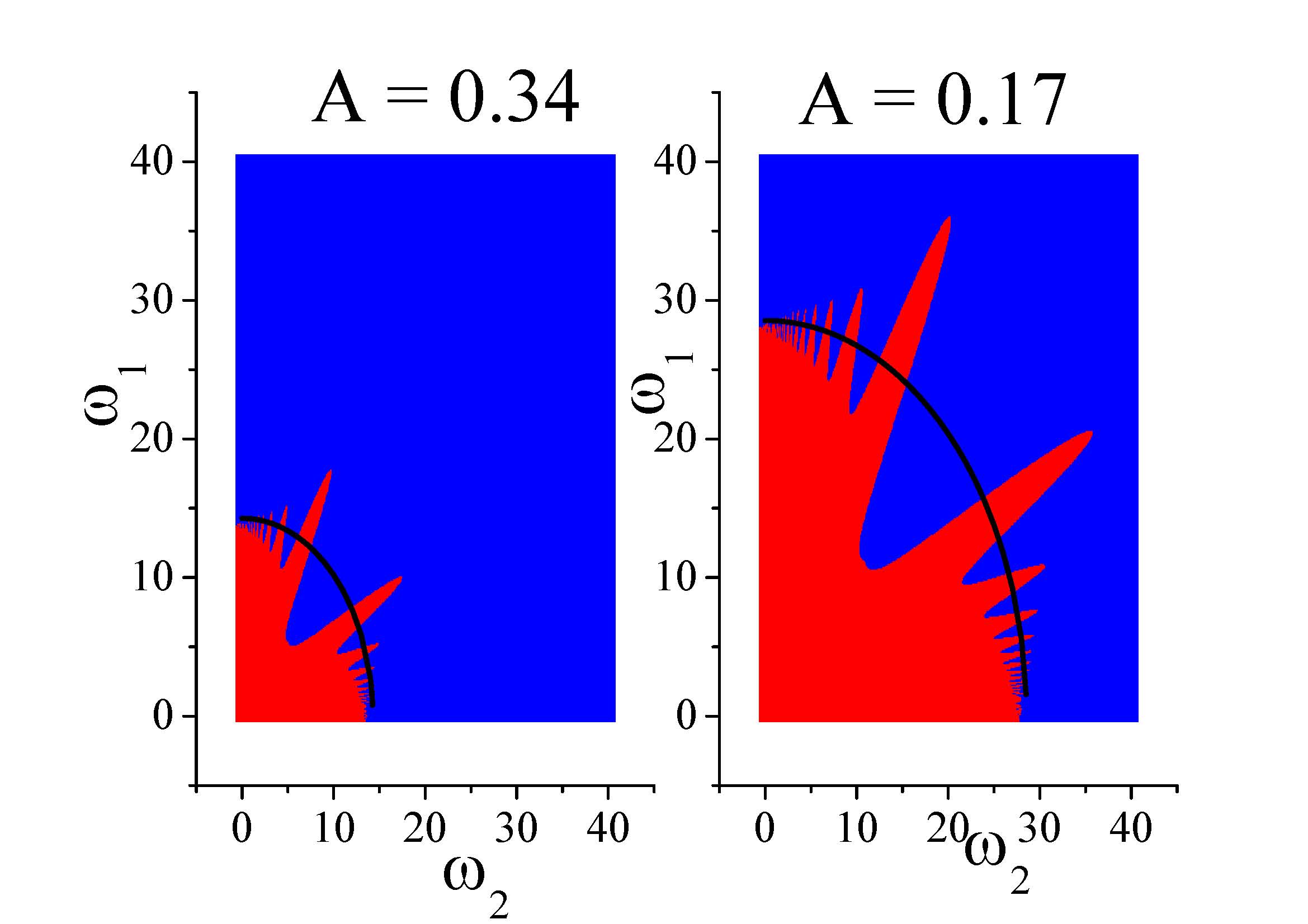}
\end{center}
\caption{Potential effective for $N=2$ according to Eq. \protect\ref{Eq.
Effective_potential_N=2}. We show that for different amplitudes we have the
same behavior which shows that effective potential as case $N=1$ has a
limitation since numerically we have very different diagrams considering $%
A=0.17$ and $A=0.34.$}
\label{Fig:effective_potential_different_amplitudes}
\end{figure}

It is clear from these results that the stability diagrams obtained from the
effective potential approximation depends strongly on the choice of $T$.
From all the numerical simulations, the best option is the choice $T=\max
(T_{1},T_{2})$. Another limitation of Eq. \ref{Eq. Effective_potential_N=2}
is shown in Fig. \ref{Fig:effective_potential_different_amplitudes}. Here,
for different amplitudes we have the same stabilization pattern, except by
the fact that the pattern is rescaled with $A$. On the contrary, however,
simulations show that $A=0.17$ and $A=0.34$ (please see again Figs. \ref%
{Fig:Neq2_A=0,17_diff_initial_angles} and \ref%
{Fig:Neq2_A=0,34_diff_initial_angles} have completely different diagrams
than ones which are shown in the Fig. \ref%
{Fig:effective_potential_different_amplitudes}. Just as in the case $N=1$,
we must also have an upper limit for the amplitude $A$, however a formulas
like Eq. \ref{Eq.parametric_ressonance} is beyond of our expectations.

Now let us studying the stochastic stabilization considering $N>2$. Dettman,
keating and Prado \cite{Prado2004} studied this problem in the context of
stochastic stabilization of chaos. And they showed not using our pendulum
inverted equation: $\frac{d^{2}\theta }{dt^{2}}=\frac{g}{l}\left( 1-\frac{A}{%
g}\sum \omega _{i}^{2}\cos (\omega _{i}t)\right) \sin \theta $ but so $\frac{%
d^{2}\theta }{dt^{2}}=\left( 1-A\sum_{i=1}^{N}\sin (\omega _{i}t+\varphi
_{i})\right) \theta $ that $\theta $ pendulum should be stabilized. Indeed
for this particular equation this indeed occurs. We tested this equation
with parameters used in this paper: $A=38$, by changing sin by cos (to bring
even more proximity with our case) and moreover making $\varphi _{1}=\varphi
_{2}=...=0$ which they did not used but which bring even more to similarity
with our case. So we also perform 100 frequencies chosen at random from
[120,\ 600] and we do stabilize. However this means to make $g=l=1$ in our
case which is not real parameters for our problem. By making $g=l=1$ we also
numerically stabilize $\theta $ even considering $\omega _{i}^{2}$ term in
the sum which does not appear in \cite{Prado2004}, but the same does not
ocurrs with real values in our case ($l=1.2$ m and $g=9.81$ m/s$^{2}$). The
general arbitrary case $\omega _{0}^{2}=g/l$ with the presence of term $%
\omega _{i}^{2}$ deserves an special attention and the problem is being
studied by the authors in another contribution (see \cite{SilvaPrado2016})

This negative case leads to look the problem in a alternative point of view
as we presented in section \ref{Section:Perturbative_Analysis}.

\begin{figure}[th]
\begin{center}
\includegraphics[width=\columnwidth]{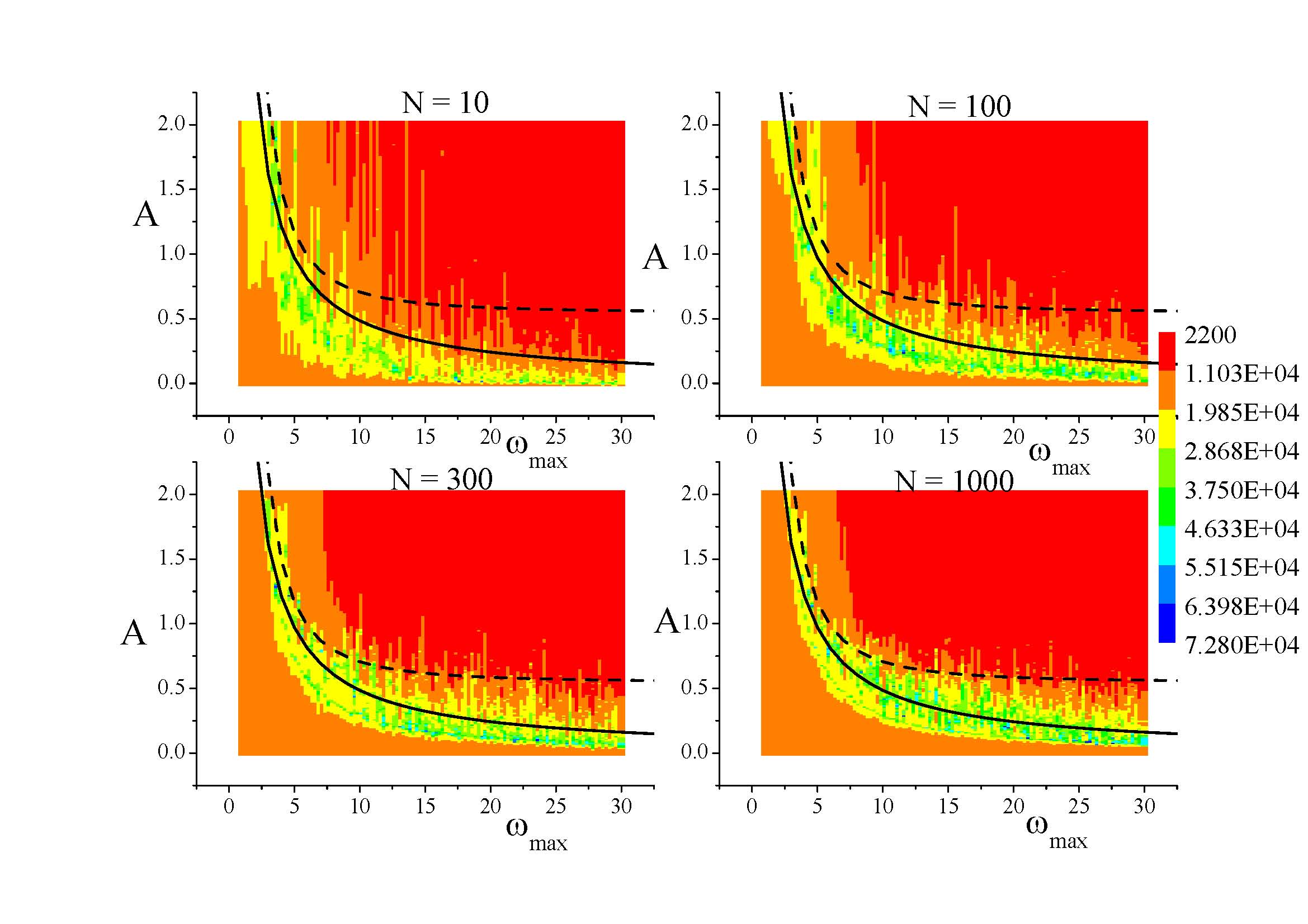}
\end{center}
\caption{Diagram $a\times \protect\omega _{\max }$. For each pair the point
corresponds to survival time (necessary time for the pendulum loses its
stability). We can observe a better agreement with region determined by
bounds as $N$ enlarges which shows that case $N=1$ has a capture according
to amplitude normalization. }
\label{Fig:Survival_time_N_infinity_changes_to_N=1}
\end{figure}

For that we consider the case for $N$ large considering a normalization for
amplitude: $A(t)=Ct$, with $CN=a\omega _{\max }$ onde $\omega _{1}$, $\omega
_{2}$,..., $\omega _{N}$ are randomly chosen in interval [$0$,$\omega _{\max
}$]. Such choice as previously reported must captures the case $N=1$ at
least for survival time. So we perform simulations (Procedure III - Table %
\ref{Table:procedure_3}) that performs a random formula with $N$ cosines. So
we build diagrams $a$ versus $\omega _{\max }$ for survival time (time that
pendulum remains stable according the established condition) which can be
observed in Fig. \ref{Fig:Survival_time_N_infinity_changes_to_N=1}.

After this interesting phenomena that brings $N$ large for $N=1$, we
concentrate our ideas for an interesting optimization process related to
stability of pendulum. What the best $N$ for a stabilization of inverted
pendulum. This question when performed so free seems to be no interesting.
However, the question is, if we would consider an ensemble of formulas by
randomly chosen $\omega _{1},...,\omega _{N}$ and $A_{1},...,A_{N}$ in the
intervals [$\omega _{\min }=0,\omega _{\max }]$ and $[A_{\min }=0,A_{\max }]$
by repeating $N_{run}=2000$ different formulas and for each choise we
observe the stability or not of the pendulum by calculating with this sample
a survival probability. We used our Procedure 4: table \ref%
{Table:Procedure_4} to calculate such probability. The Fig. \ref%
{Fig:survival probability} shows the survival probability in different
situations.

\begin{figure}[th]
\begin{center}
\includegraphics[width=0.6\columnwidth]{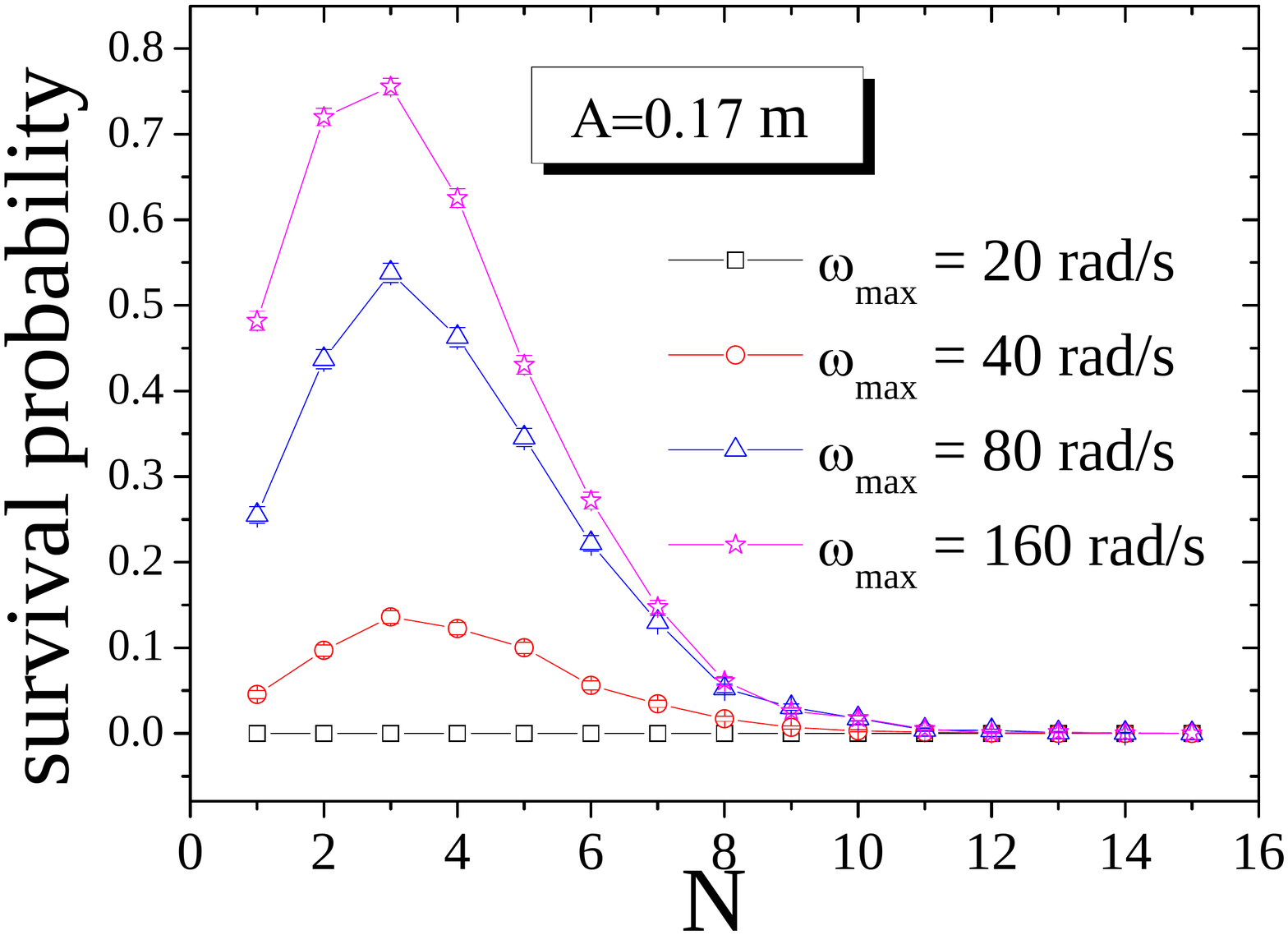} 
\includegraphics[width=0.6%
\columnwidth]{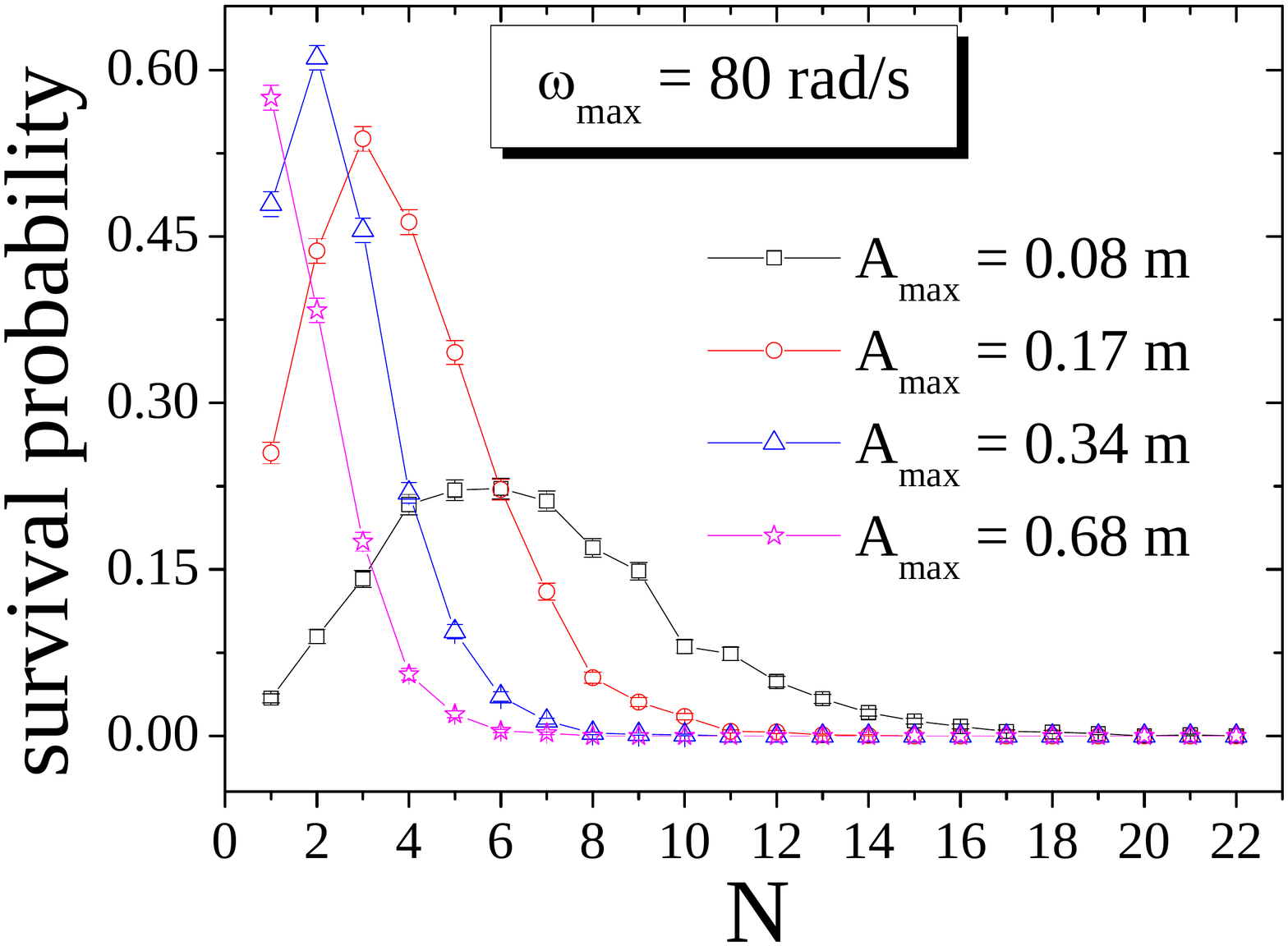} 
\includegraphics[width=0.6%
\columnwidth]{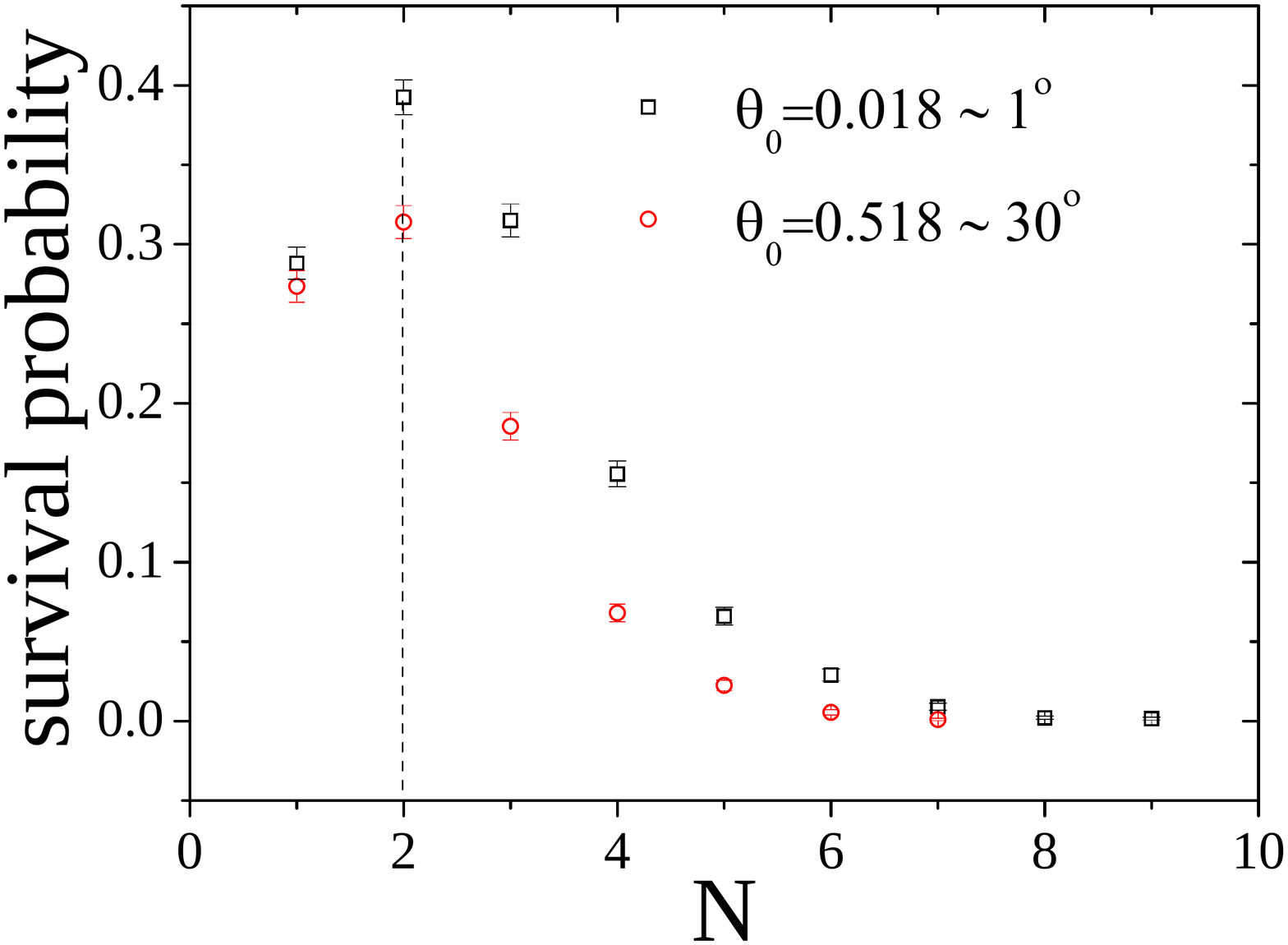}
\end{center}
\caption{Survival probability according to Procedure IV.}
\label{Fig:survival probability}
\end{figure}

The upper plot in this figure by keeping $A_{\max }=0.17$m and we plot the
probability for different values of $\omega _{\max }$. The different
frequencies does not change the $N_{\max }$(value that maximizes the
survival probability). However the midle figure, shows that keeping $\omega
_{\max }$ fixed and by plotting the survival probabilities for different
values $A_{\max }$. In this case we change the $N_{\max }$. But it is
important to notice that different initial angles does not change $N_{\max }$
as reported in the lower plot in same Fig. \ref{Fig:survival probability}.

\section{Conclusions}

\label{Section:Conclusions}

In this paper we have detailed the study of inverted pendulum under a
parametric excitation in its basis which is described by a superposition of $%
N$ cosines. In case $N=1$ we explore diagrams $A$ $\times \ \omega $. We
show that depending of initial conditions the effective potential method
diverges from the numerical simulations, which also occurs in $N=2$ that
presents a interesting diagram where stability regions are alternated with
no stability ones in a fractal structure. The diagonal $\omega _{1}=\omega
_{2}=\omega $ has a important hole in the stability due to the known bating
problem in waves. Although the effective potential method is extended
for arbitrary $N$ is extended for arbitrary cases its usuality depends on
choice of a common period existence and its utility has several limitations
by showing the necessity of Runge Kutta integrations of equations which in
this paper is separated in four procedures used in each part of our
manuscript and showed in details.

For $N>2$ we perform two kind of analysis: a) a discussion about stochastic
stabilization and b) optimization of survival probability of pendulum. In
this first part (a), starting from hypothesis that inverted pendulum with
real parameters cannot be stabilized for $N$ large (this not occurs with
suitable choice of parameters) we show that problem for $N$ large can be
reduced for $N=1$ if we look for survival time, i.e., the properties of
survival time diagram when via a suitable scale in the amplitude are
preserved when compared with regular, $N=1$ diagram. In second part (b) we
choose randomly choose amplitudes and frequencies in ranges and we calculate
the survival probability of pendulum in order to observe the optimal $N$
that maximizes such probability. We have two important three important
conclusions here by observing our numerical studies : a) By fixing the upper
limit of the frequencies chosen and changing the amplitude, $N_{opt}$
depends on amplitude b) By fixing the upper limit of the amplitude and
changing frequency, $N_{opt}$ remains the same. The initial angles seems to
be does not change $N_{opt}$.

\textbf{Acknowledgments --} This research was partially supported by the
Conselho Nacional de Desenvolvimento Cientifico e Tecnologico (CNPq), under
the grant 11862/2012-8. The authors would like to thank Prof. L.G. Brunet
(IF-UFRGS) for kindly providing the computational resources from Clustered
Computing (ada.if.ufrgs.br) for this work.(2014)

\end{document}